\documentclass{aa}  

\usepackage{graphicx}
\usepackage{txfonts}
\usepackage{hyperref}
\begin{document}
\title{Apsidal motion and proximity effects in the massive binary BD+60$^{\circ}$\,497\thanks{Based on optical spectra collected at Observatoire de Haute Provence (France) and with the TIGRE telescope (La Luz, Mexico).}}
\author{G.\ Rauw\inst{1} \and P.A.\ {Ko{\l}aczek-Szyma{\'n}ski}\inst{1,2} \and \ Y.\ Naz\'e\inst{1}\thanks{F.R.S.-FNRS Senior Research Associate} \and L.\ Nys\inst{1}}
\institute{Space sciences, Technologies and Astrophysics Research (STAR) Institute, Universit\'e de Li\`ege, All\'ee du 6 Ao\^ut, 19c, B\^at B5c, 4000 Li\`ege, Belgium
\and  University of Wroc{\l}aw, Faculty of Physics and Astronomy, Astronomical Institute, ul. Kopernika 11, 51-622 Wroc{\l}aw, Poland \\ \email{g.rauw@uliege.be}}
\date{}

  \abstract
      {The eccentric short-period O-star binary BD+60$^{\circ}$\,497 is an interesting laboratory in which to study tidal interactions in massive binary systems, notably via the detection and characterisation of apsidal motion.}{The rate of apsidal motion in such systems can help constrain their age and provide insight into the degree of mass concentration in the interior of massive stars.}
      {We used spectroscopic data collected over two decades to reconstruct the individual spectra of the stars and to establish their epoch-dependent radial velocities. An orbital solution, explicitly accounting for apsidal motion was adjusted to the data. Space-borne photometric time series were analysed with Fourier methods and with binary models.}{We derived a rate of apsidal motion of $\dot{\omega} = (6.15^{+1.05}_{-1.65})^{\circ}$\,yr$^{-1}$, which suggests an age of $4.13^{+0.42}_{-1.37}$\,Myr. The disentangled spectra unveiled a curious change in the spectral properties of the secondary star between the epochs 2002-2003 and 2018-2022, with the secondary spectrum appearing to be of an earlier spectral type over recent years. Photometric data show variability at the $\sim 6$\,mmag level on the period of the binary system, which is hard to explain in terms of proximity effects.}{Whilst the rate of apsidal motion agrees well with theoretical expectations, the changes in the reconstructed secondary spectrum hint at a highly non-uniform surface temperature distribution for this star. Different effects are discussed that could contribute to the photometric variations. The current most-likely explanation is a mix of proximity effects and tidally excited oscillations.}
\maketitle
\section{Introduction}
Over the past two decades, binarity and higher multiplicity has been established as a common property of massive stars \citep[e.g.][and references therein]{Off23}. Whilst this multiplicity leads to more complex evolutionary paths \citep[e.g.][]{Che24}, it nonetheless offers new opportunities for the observational determination of the physical properties of massive stars. The most obvious example is the study of eclipsing spectroscopic binaries, which provides the least model-dependent way to determine the masses and radii of stars, as it simply relies on Newton's law of gravity and geometric models based on first principles \citep[e.g.][]{Mah20}.  

Furthermore, tidal interactions in binary systems can help probe the internal structure of stars. Among the most important consequences of the tidal interactions in short-period eccentric stellar binary systems is the Newtonian apsidal motion \citep[e.g.][]{Har14,Sch16,Zas20,Ros20a,Cla21}. Apsidal motion consists of a slow secular change of the argument of periastron, $\omega$, triggered by the non-spherical gravitational field of tidally distorted stars. The rate of this apsidal motion, $\dot{\omega}$, depends on the internal structure constant of order two ($k_2$; e.g.\ \citealt{Sha85}). This latter quantity can be determined from theoretical stellar structure models via the resolution of the Radau differential equation \citep[e.g.][]{RRN16,Ros20b}. Hence, measuring $\dot{\omega}$ provides an efficient way to constrain the internal structure of stars \citep{Cla10,Ros20b}.

Tidal interactions also have other observational consequences. In close eccentric binary systems, the tides are dynamical, as they vary over the orbital cycle. This can trigger tidally excited oscillations (TEOs), provided there exists a resonance between the frequencies of the eigenmodes of the components and integer multiples of the orbital frequency \citep[see e.g.][]{Wil03,Wil07,Kol23,Fel25}. TEOs thus manifest themselves in photometric light curves as pulsational modulations of the brightness at harmonics of the orbital frequency. Moreover, the phase-dependent tidal deformation of the stars along with the phase-modulated light reflection and mutual heating effects lead to additional photometric variations such as the so-called heartbeat variation near periastron \citep{Pab17,Eng20,Kol21}.

Short-period ($P_{\rm orb} \lesssim 20$\,d) eccentric binaries are thus expected to display a variety of proximity effects in their light curves and spectra. In this context, we revisit the short-period eccentric O-star binary BD+60$^{\circ}$\,497 \citep{Rau04, Hil06,Rau16} in the young open cluster IC\,1805. This non-eclipsing binary system hosts an O6.5\,V((f)) primary and a late-O secondary on a $3.96$\,d orbit of moderate eccentricity \citep[$e \sim 0.16$;][]{Rau16}. These properties make BD+60$^{\circ}$\,497 a promising target to search for apsidal motion and other proximity effects.  

\section{Observations and data reduction}
\subsection{Spectroscopy}
Spectroscopic observations of BD+60$^{\circ}$\,497 were collected during five observing campaigns at the Observatoire de Haute Provence (OHP) in France using the 1.52\,m telescope equipped with the Aur\'elie spectrograph \citep{Gillet}. These observations covered the spectral range between 4450 and 4890\,\AA\ at a resolving power of $\sim$ 10\,000. For the data collected in September 2002, October 2003, and August 2018, the detector was an EEV CCD with $2048 \times 1024$ pixels of size 13.5\,$\mu$m squared. In August 2019 and September 2022, the detector was replaced by an Andor CCD camera with $2048 \times 512$ pixels of the same size. The data from the 2002 and 2003 campaigns were previously described by \citet{Rau04} and used by \citet{Rau16}. Integration times of individual exposures ranged between 30\,min and 45\,min, depending on the weather conditions. The OHP spectra were reduced using version 21FEBpl\,1.0 of the {\sc midas} software developed at the European Southern Observatory.  

In 2014 and 2018, we collected a series of observations with the HEROS echelle spectrograph on the robotic 1.2\,m TIGRE telescope \citep{Schmitt,Gon22} installed at La Luz Observatory near Guanajuato (Mexico). TIGRE/HEROS spectra have a resolving power of 20\,000 over the spectral range from 3760 -- 8700\,\AA\ with a small gap around 5600\,\AA. Due to an instrumental issue, the data in the blue channel of the 2018 campaign could not be used. The data reduction was performed with the dedicated HEROS pipeline \citep{Mittag,Schmitt}. Telluric absorptions in the spectral regions around the He\,{\sc i} $\lambda$\,5876, H$\alpha$ and He\,{\sc i} $\lambda$\,7065 lines were corrected by means of the {\tt telluric} routine within {\sc iraf} and using the atlas of telluric lines of \citet{Hinkle}.

The journal of the observations used in this work for spectral disentangling and radial velocity (RV) analysis is provided in Table\,\ref{Journal}. Our spectra were continuum normalised using {\sc midas}, adopting the best-fit spline functions adjusted to a set of continuum windows. Preliminary RVs of both stars were determined by adjusting double Gaussian profiles to the He\,{\sc i} $\lambda\lambda$\,4471 and 4713 lines. These estimated RVs were then used as input values to perform spectral disentangling. For this purpose, we used our implementation of the shift-and-add spectral disentangling method described by \citet{Gon06}. This method takes advantage of the Doppler shifts at different orbital phases to iteratively reconstruct the individual spectra of the components of the binary system and simultaneously improves our estimates of the individual RVs at each epoch of observation \citep[e.g.][]{Rau16}. At each iteration, to establish the RVs of a given component at a specific date, the Doppler-shifted reconstructed spectrum of the companion is first subtracted from the spectrum observed on that date. The residuals are then cross-correlated with a synthetic TLUSTY spectrum \citep{Lan03}, and the updated RVs are evaluated by adjusting a parabola to the peak of the cross-correlation function \citep{Ver99}. For the primary and secondary stars of BD+60$^{\circ}$\,497, we used TLUSTY spectra respectively with effective temperatures $T_{\rm eff} = 37.5$\,kK and $T_{\rm eff} = 32.5$\,kK. Both template spectra assumed a surface gravity $\log{g} = 4.0$ and were broadened to a projected rotational velocity of $v\,\sin{i} = 150$\,km\,s$^{-1}$.

In the blue part of the spectrum, we disentangled the spectral ranges from 4450 to 4600\,\AA, from 4600 to 4780\,\AA, and from 4780 to 4890\,\AA. Each of these domains contains at least one strong spectral line. We iterated our disentangling routine until the maximum absolute value of the RV correction between two consecutive iterations became less than 1.0\,km\,s$^{-1}$. For most wavelength ranges convergence was achieved within fewer than 100 iterations. 

\subsection{Photometry}
The Transiting Exoplanet Survey Satellite \citep[{\it TESS};][]{TESS} observed BD+60$^{\circ}$\,497 during four sectors (18 in November 2019, 58 in November 2022, and 85 and 86 from the end of October 2024 until mid December 2024). Broadband photometry over the 6000\,\AA\, to 1\,$\mu$m bandpass was collected at a 2\,min cadence. We retrieved the light curves, processed with the {\it TESS} pipeline \citep{Jen16}, from the Mikulski Archive for Space Telescopes (MAST) portal.\footnote{http://mast.stsci.edu/} In our analysis, we used the pre-search data conditioned (PDC) photometry obtained after removing trends correlated with systematic spacecraft or instrumental effects and correcting for crowding (if any). We filtered the data to keep only those with a quality flag of zero and converted the PDC fluxes into magnitudes.

Because the pixels of the {\it TESS} CCD detectors encompass an area of 21\arcsec $\times$ 21\arcsec on the sky and since the aperture photometry is extracted over several pixels, we checked the {\it Gaia} data release 3 catalogue \citep[DR3,][]{DR3} for possible contaminators within a 1\arcmin\ radius of our target. The catalogue lists 71 objects, all of which are significantly fainter than our target. The brightest neighbouring source is 6.3\,mag fainter than BD+60$^{\circ}$\,497. Hence, the {\it TESS} photometry does not suffer from significant contamination by neighbouring sources.

\section{Spectroscopy}
\subsection{Reconstructed spectra \label{recspec}}
\begin{figure}
  \includegraphics[angle=0, width=8.5cm]{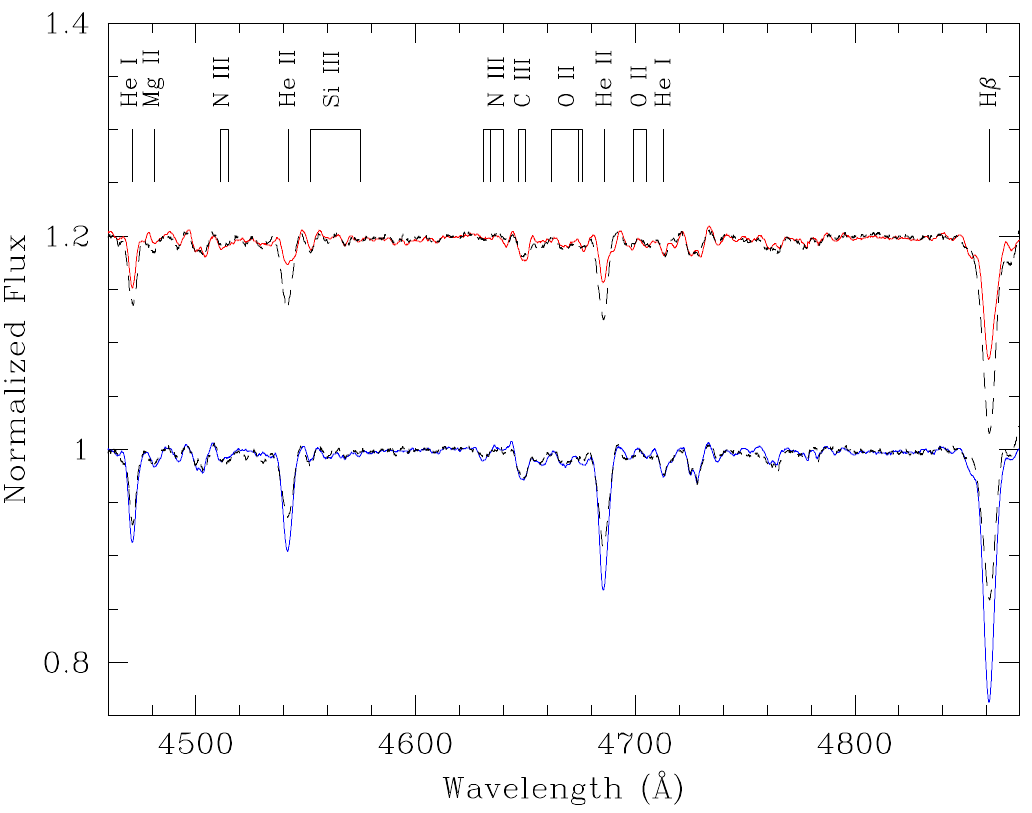}
  \caption{Disentangled Aur\'elie spectra of BD+60$^{\circ}$\,497. Reconstructed spectra from the 2002-2003 epoch are shown by the blue continuous line for the primary, and the red continuous line for the secondary (shifted vertically by 0.2 units). The black dashed lines represent the result of the disentangling for the 2018-2022 data. The spectra are displayed normalised to the global continuum of the system. The broad double-minimum features around 4502, 4726, 4763, and 4780\AA\ are artefacts due to the disentangling of the stationary diffuse interstellar bands at those wavelengths. \label{specOHP}}
\end{figure}

Spectral disentangling was applied independently to the full set of Aur\'elie data collected in September 2002 and October 2003 (16 spectra) on the one hand, and all Aur\'elie data collected after August 2018 (28 spectra) on the other hand. The reconstructed spectra are shown in Fig.\,\ref{specOHP}. We found significant differences in the reconstructed spectra of the 2002-2003 and 2018-2022 epochs which are not due to differences in spectral resolution as all data were taken with the same instrumentation. These differences concern most of all the He\,{\sc ii} $\lambda\lambda$\,4542 and 4686 lines. In 2018-2022, these lines were considerably stronger in the secondary spectrum than in 2002-2003. For completeness, we note that the disentangled blue spectra of the 2014 TIGRE/HEROS campaign are in excellent agreement with those of the 2002-2003 Aur\'elie campaign. 

Such differences might stem from mutual heating effects and a peculiar sampling of the orbital cycle. Indeed, \citet{Rau04} noted that the visibility of the secondary spectral lines changed with orbital phase. The 2002-2003 data contain mostly data taken at quadrature orbital phases, whilst 35\% of the data from 2018-2022 were taken near conjunction phases. To check whether this difference in sampling biases our reconstruction, we repeated the spectral disentangling restricting ourselves to spectra with a RV separation between the two components $|RV_p - RV_s| \geq 200$\,km\,s$^{-1}$. In this way, we retained 14 spectra (out of 16) for the 2002-2003 epoch and 18 spectra (out of 28) for the 2018-2022 epoch. Despite the different samplings, the reconstructed spectra in the second approach were essentially unchanged compared to our results with the full datasets of each epoch.  

\begin{figure}
  \includegraphics[angle=0, width=8.5cm]{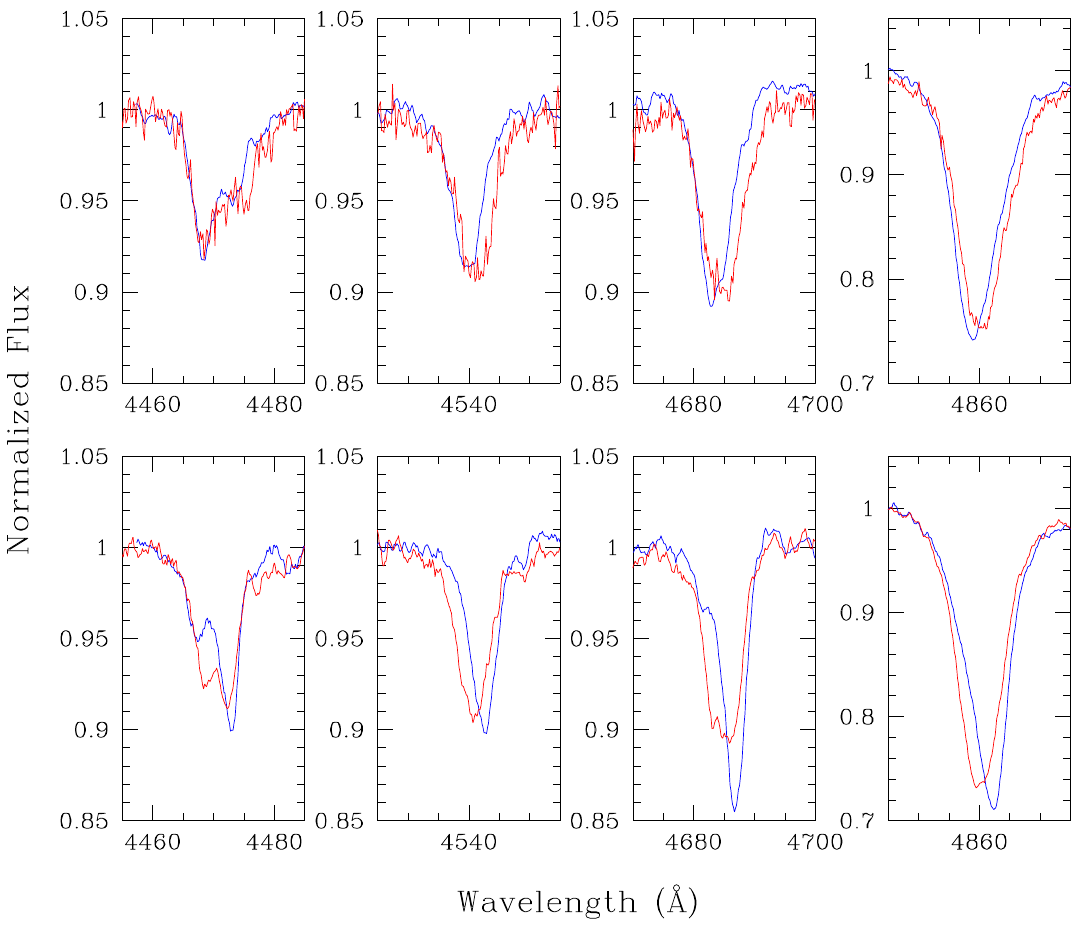}
  \caption{Comparison between the strongest lines in the Aur\'elie spectra as observed in 2002-2003 (blue) and 2018-2022 (red). The lines are, from left to right He\,{\sc i} $\lambda$\,4471, He\,{\sc ii} $\lambda$\,4542, He\,{\sc ii} $\lambda$\,4686, and H$\beta$. The top panel illustrates the spectra observed close to the quadrature with the primary moving towards the observer, whilst the bottom panel illustrates the comparison for phases near the opposite quadrature. Spectra in the top panel were taken on HJD\,2452928.641 (blue) and HJD\,2459853.5509 (red), that is respectively at phases 0.99 ($\omega = 150^{\circ}$, position angle $236^{\circ}$) and 0.78 ($\omega = 261^{\circ}$, position angle $254^{\circ}$) according to our 'RV dataset 2' ephemerides in the forthcoming Table\,\ref{omegadotfit}. Spectra in the bottom panel were obtained on HJD\,2452527.555 (blue) and HJD\,2458719.6140 (red), that is respectively phases 0.70 ($\omega = 144^{\circ}$, position angle $111^{\circ}$) and 0.42 ($\omega = 243^{\circ}$, position angle $130^{\circ}$) according to the same ephemerides.\label{montage}}
\end{figure}

Figure\,\ref{montage} compares observed spectra (i.e. prior to any disentangling) collected near quadrature phases during the 2002-2003 and 2018-2022 epochs. One can clearly see a difference in relative strength of the lines of the two stars between those epochs. This is especially apparent for the quadrature phase when the primary star is receding. At those phases, the secondary absorptions are nearly of equal strength to those of the primary in the 2018-2022 data, whereas they were considerably weaker in 2002-2003. An alternative to the mutual heating explanation could be the blend with the spectral signature of a third component. However, \citet{Mas09} did not detect any companion at an angular separation down to 0.03\arcsec and with a $V$-band magnitude difference $\Delta V \leq 3$\,mag in their speckle interferometric survey. Similarly, \citet{Mai10} did not detect a companion using lucky imaging with a separation down to 0.1\arcsec and $\Delta V \leq 8$\,mag. {\it Gaia} DR3 \citep{DR3} quotes a Renormalised Unit Weight Error (RUWE) value of 1.19 for BD+60$^{\circ}$\,497. This parameter expresses the quality of the adjustment of the astrometric data. If a single source model provides a good fit, RUWE is expected to be around 1.0. In our case, the RUWE value does not provide evidence of a non-single source.

We attempted to disentangle the whole set of Aur\'elie spectra accounting for the possible presence of a third star. However, as could be expected from the absence of a clear signature of a third star, these attempts did not result in a meaningful reconstruction of the spectra. 

\citet{Rau04} classified BD+60$^{\circ}$\,497 as O6.5\,V((f)) + O8.5-9.5\,V((f)) based on observations taken around opposite quadrature phases. \citet{Sot11} classified BD+60$^{\circ}$\,497 as O6.5\,V((f)) + O8/B0\,V. \citet{Rau16} used disentangled spectra to infer an O6.5\,V((f)) + O8.5\,V classification. Comparison of our disentangled spectra for the 2002-2003 epoch with the standard classification spectra of \citet{Sot11} indicates a primary spectrum consistent with an O6.5-7\,V classification, except for the unusually strong Si\,{\sc iv}\,4631,  C\,{\sc iii} $\lambda\lambda$\,4647-50-51, O\,{\sc ii} $\lambda\lambda$\,4662-72 and O\,{\sc ii} $\lambda\lambda$\,4699-4705 absorptions which would be more consistent with a slightly later (O7.5) spectral type. On the other hand, the disentangled spectra from the 2018-2022 epoch yield a primary spectrum with weaker H\,{\sc i}, He\,{\sc i}, and He\,{\sc ii} absorption lines. The He\,{\sc i}/He\,{\sc ii} ratio remains consistent with a spectral type around O7. These results are confirmed when considering the \citet{Con71} and \citet{Mat88,Mat89} criteria based on the equivalent width (EW) ratios of the He\,{\sc i} $\lambda$\,4471 and He\,{\sc ii} $\lambda$\,4542 lines. At both epochs, we obtain ratios at the boundary between O6.5 and O7. Comparing with the grid of synthetic TLUSTY O-star spectra \citep{Lan03}, we obtain a good match for the primary with an effective temperature of 37.5\,kK and a fractional optical luminosity of 65\% for the 2002-2003 epoch and of 50\% for the 2018-2022 epoch.  

The differences in the spectral appearance between the two epochs are more pronounced for the disentangled secondary spectrum. During the 2002-2003 epoch, the secondary spectrum was consistent with spectral type O8.5\,V, but the strength of H\,{\sc i}, He\,{\sc i} and He\,{\sc ii} lines strongly increased in the 2018-2022 epoch. The intensity ratio of the helium lines of that epoch points towards an O7\,V spectral type. Quite interestingly, the metal lines in the disentangled spectra of both stars do not undergo such strong changes. Comparing again with the grid of TLUSTY spectra, we obtain a good match for the secondary in 2002-2003 with $T_{\rm eff} \simeq 32.5$\,kK and a fractional optical luminosity of 35\%. To match the 2018-2022 spectrum we need instead a model with $T_{\rm eff} \simeq 36.25$\,kK and contributing about 50\% of the optical flux.  

\subsection{Rotational velocity and optical brightness}
We applied the Fourier transform method \citep{Sim07} to the Si\,{\sc iii} $\lambda$\,4552 line in the disentangled primary and secondary spectra to determine the projected rotational velocity, $v\,\sin{i}$. The best-fit rotational broadening functions correspond to $v\,\sin{i} = 130 \pm 20$\,km\,s$^{-1}$ for the primary and  $110 \pm 20$\,km\,s$^{-1}$ for the secondary. The quoted uncertainties correspond to the dispersions of the measurements between the 2002-2003 and 2018-2022 epochs; the $v\,\sin{i}$ values of both stars were found to be higher in 2018-2022. 

The $V$ magnitude and $B-V$ colour of BD+60$^{\circ}$\,497 were given as 8.773 and 0.566 by \citet{Sun95}. This is in good agreement with the mean values of the measurements compiled by \citet{Ree03}: $V = 8.80 \pm 0.03$ and $B-V = 0.565 \pm 0.011$. Adopting $(B-V)_0 = -0.27$ \citep{Mar06}, we infer a colour excess of $0.835 \pm 0.011$. \citet{Sun17} derived a total-to-selective extinction ratio towards IC\,1805 of $R_V = 3.05 \pm 0.06$. This yields $A_V = 2.547 \pm 0.060$. \citet{Mai18} inferred a somewhat higher $R_V = 3.182 \pm 0.056$ and thus a larger $A_V = 2.692 \pm 0.018$, which would imply a 13\% increase of the luminosity of the system compared to the \citet{Sun17} extinction ratio. Considering nine O-type stars in IC\,1805, the mean geometric distance inferred from the study of \citet{Bai21} based on {\it Gaia} EDR3 is $(2021 \pm 186)$\,pc. For the specific case of BD+60$^{\circ}$\,497, \citet{Bai21} quoted a somewhat larger geometric distance $(2440^{+128}_{-114})$\,pc. Whilst the astrometric parameters of BD+60$^{\circ}$\,497 position the system outside the main locus of O-type stars in IC\,1805, it still lies inside the locus of OB stars (considering B-type stars up to spectral type B3, see Appendix\,\ref{AppB}) which are presumably members of IC\,1805. Adopting the mean IC\,1805 distance, and the extinction ratio of \citet{Sun17}, we derive an absolute magnitude of the binary $M_V = -5.30 \pm 0.21$ that corresponds nearly exactly to the combination of the brightness of an O6.5-7\,V and an O8\,V star as listed by \citet{Mar05}. If we use instead the larger distance estimated for BD+60$^{\circ}$\,497, we obtain $M_V = -5.71 \pm 0.13$, that corresponds to a 45\% higher luminosity.  

Comparing the EWs of the He\,{\sc i} $\lambda$\,4471 and He\,{\sc ii} $\lambda\lambda$\,4542, 4686 lines measured on our disentangled spectra with the EWs computed from TLUSTY synthetic spectra for stars of same spectral type, we can estimate the fractional $V$-band luminosities of both stars as $0.60 \pm 0.10$ for the primary and $0.40 \pm 0.17$ for the secondary. These results imply that any putative third light contribution must be of low level (less than 20\% of the total light).

\subsection{Orbital solution and apsidal motion \label{apsidal}}
Literature RVs of our target were compiled by \citet{Mer84}\footnote{Available as catalogue III/97A at the Centre de Donn\'ees Stellaires, https://vizier.cds.unistra.fr/viz-bin/VizieR-3?-source=III/97A.}. The original data came from \citet{Hay32} and \citet{Und67}. With the exception of one measurement by \citet{Und67}, these data did not resolve the SB2 nature of the system. This could bias the RVs. Combining these literature RVs with the measurements provided by \citet{Hil06} and our own RVs obtained via spectral disentangling of the Aur\'elie data, for dates when the RV separation between the two stars was larger than 200\,km\,s$^{-1}$, results in a set of 62 primary RV points spread over nearly a century. The vast majority of these points (57 out of 62) were acquired within the last 20 years.

To establish an orbital solution for the primary that accounts for the effects of apsidal motion, we used the method of \citet{RRN16} and \citet{Ros20a}, where the primary RV data are adjusted by the mathematical expression
\begin{equation}
  RV_p(t) = \gamma + K_p\,\left[\cos(\phi(t) + \omega(t)) + e\,\cos(\omega(t))\right],
\label{RVomegadot}  
\end{equation}  
where  $\gamma$ is the systemic velocity, $K_p$ is the semi-amplitude of the primary RV curve, $\phi(t)$ is the true anomaly (computed via Kepler's equation), $e$ is the eccentricity, and $\omega(t)$ is the longitude of periastron, which explicitly accounts for apsidal motion:
\begin{equation}
  \omega(t) = \omega_0 + \dot{\omega}\,(t - t_0),
\end{equation}
with $\omega_0 = \omega(t_0)$ and $t_0$ as a reference time of periastron passage. The rate of apsidal motion, $\dot{\omega}$, hence becomes an explicit model parameter, along with $K_p$, $e$, the anomalistic period $P_{\rm anom}$ (that is the period between two consecutive periastron passages), $\omega_0$ and $t_0$.

Our method consists then in scanning a six-dimensional parameter space and searching for the combination of parameters that best adjusts the observed $RV_p(t)$ values. Because the secondary RVs are subject to larger uncertainties, we restricted the apsidal motion analysis to the primary RVs. After some preliminary explorations, we built a grid of models with $K_p$ varying from 137.5 to 160.5\,km\,s$^{-1}$, $P_{\rm anom}$ between 3.9587\,d and 3.9614\,d, $t_0$ between HJD\,2456681.8 and HJD\,2456683.8, $e$ between 0.0 and 0.30, $\omega_0$ between 130 and $280^{\circ}$, and $\dot{\omega}$ between 0 and $15^{\circ}$\,yr$^{-1}$. The steps were respectively $0.5$\,km\,s$^{-1}$, $6 \times 10^{-5}$\,d, 0.1\,d, 0.01, $2^{\circ}$, and $0.15^{\circ}$\,yr$^{-1}$.

\begin{figure}
  \includegraphics[angle=0, width=8.5cm]{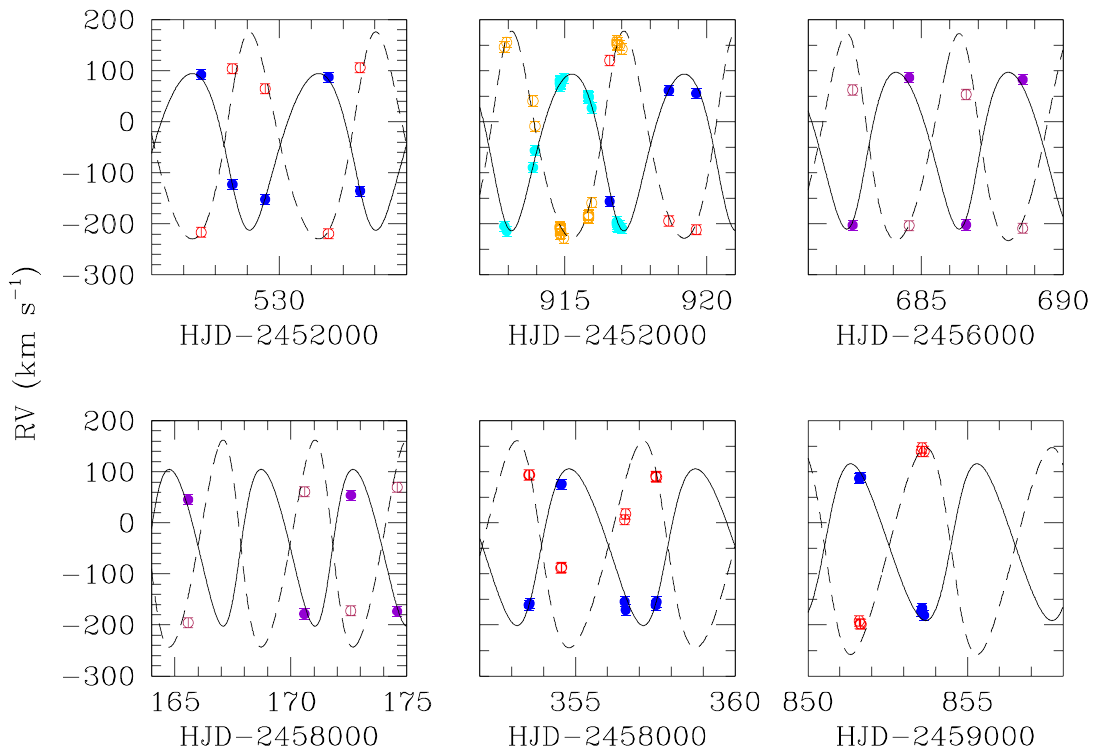}
  \caption{Comparison between the best-fit model for the primary RVs (black solid curve) computed with relation \ref{RVomegadot} for dataset 2. Cyan, blue, and violet filled symbols stand respectively for primary RVs from \citet{Hil06} and our disentangling of the Aur\'elie and HEROS spectra. The orange, red, and maroon open symbols show the secondary RVs (not used in the fit) respectively from \citet{Hil06}, our Aur\'elie and HEROS data, whilst the dashed line illustrates the secondary RV solution inferred from the primary RV curve for a mass ratio $q = m_p/m_s = 1.28 \pm 0.03$.\label{RVsol}}
\end{figure}

We applied this method both to the full set of 62 primary RVs (dataset 1), keeping in mind the caveat concerning the oldest measurements, and restricting ourselves to the 57 RVs from \citet{Hil06} and our Aur\'elie RVs (dataset 2, see Fig.\,\ref{RVsol}). In both cases, we obtained the best adjustment for $K_p$ near 153\,km\,s$^{-1}$. The best-fit parameters of the two datasets overlap, with the fit of dataset 2 being of better quality (see Table\,\ref{omegadotfit}). Our determination of the rate of apsidal motion for dataset 2 yields a period of apsidal motion $U = 59^{+16}_{-10}$\,yr. The eccentricity is very much consistent with the values previously inferred by \citet{Hil06} and \citet{Rau16}. 

\begin{table}
  \caption{Best-fit orbital solutions accounting for apsidal motion.\label{omegadotfit}}
  \begin{center}
  \begin{tabular}{l c c}
    \hline
    Parameter & RV dataset 1 & RV dataset 2 \\
    \hline
    \vspace*{-3mm}\\
    $K_p$ (km\,s$^{-1}$) & $152.5^{+6.5}_{-6.5}$ & $153.5^{+4.5}_{-4.8}$\\
    \vspace*{-3mm}\\
    $P_{\rm anom}$ (d) & $3.95972^{+.00018}_{-.00017}$ & $3.95984^{+.00016}_{-.00020}$ \\
    \vspace*{-3mm}\\
    $t_0$ (HJD-2450000) & $6682.6^{+0.2}_{-0.2}$ & $6682.7^{+0.1}_{-0.2}$ \\
    \vspace*{-3mm}\\
    $e$ & $0.16^{+0.05}_{-0.06}$ & $0.15^{+0.04}_{-0.04}$ \\
    \vspace*{-3mm}\\
    $\omega_0$ ($^{\circ}$) & $210^{+24}_{-22}$ & $220^{+10}_{-20}$ \\
    \vspace*{-3mm}\\
    $\dot{\omega}$ ($^{\circ}$\,yr$^{-1}$) & $4.80^{+1.70}_{-1.05}$ & $6.15^{+1.05}_{-1.65}$ \\
    \vspace*{-3mm}\\
    \hline
    $\chi^2_{\nu}$ & 2.055 & 1.019 \\
    \hline
    $m_p\,\sin^3{i}$ (M$_{\odot}$) &  & $9.26 \pm 0.90$ \\
    $m_s\,\sin^3{i}$ (M$_{\odot}$) &  & $7.23 \pm 0.68$ \\ 
    \hline
  \end{tabular}
  \end{center}
  \tablefoot{Our best solution corresponds to the column `RV dataset 2'. The determination of the uncertainties of the parameters is described in Appendix\,\ref{Appnew} and Fig.\,\ref{cornerplot}.}
\end{table}

Finally, we used the RVs of the primary and secondary stars to adjust the mass ratio $q = m_p/m_s = 1.28 \pm 0.03$. Our value of $q$ is in excellent agreement with those of \citet{Rau04} and \citet{Hil06}, but smaller than the 1.56 value of \citet{Rau16}. With our new value of $q$, we infer minimum masses of $m_p\,\sin^3{i} = (9.26 \pm 0.90)$\,M$_{\odot}$ and $m_s\,\sin^3{i} = (7.23 \pm 0.68)$\,M$_{\odot}$. Comparing these minimum masses with the 'typical' masses taken from \citet{Mar05} for stars in the range of spectral types inferred above, we conclude that the orbital inclination is likely close to $(44.8 \pm 2.5)^{\circ}$.

\begin{figure}
  \includegraphics[angle=0, width=8.5cm]{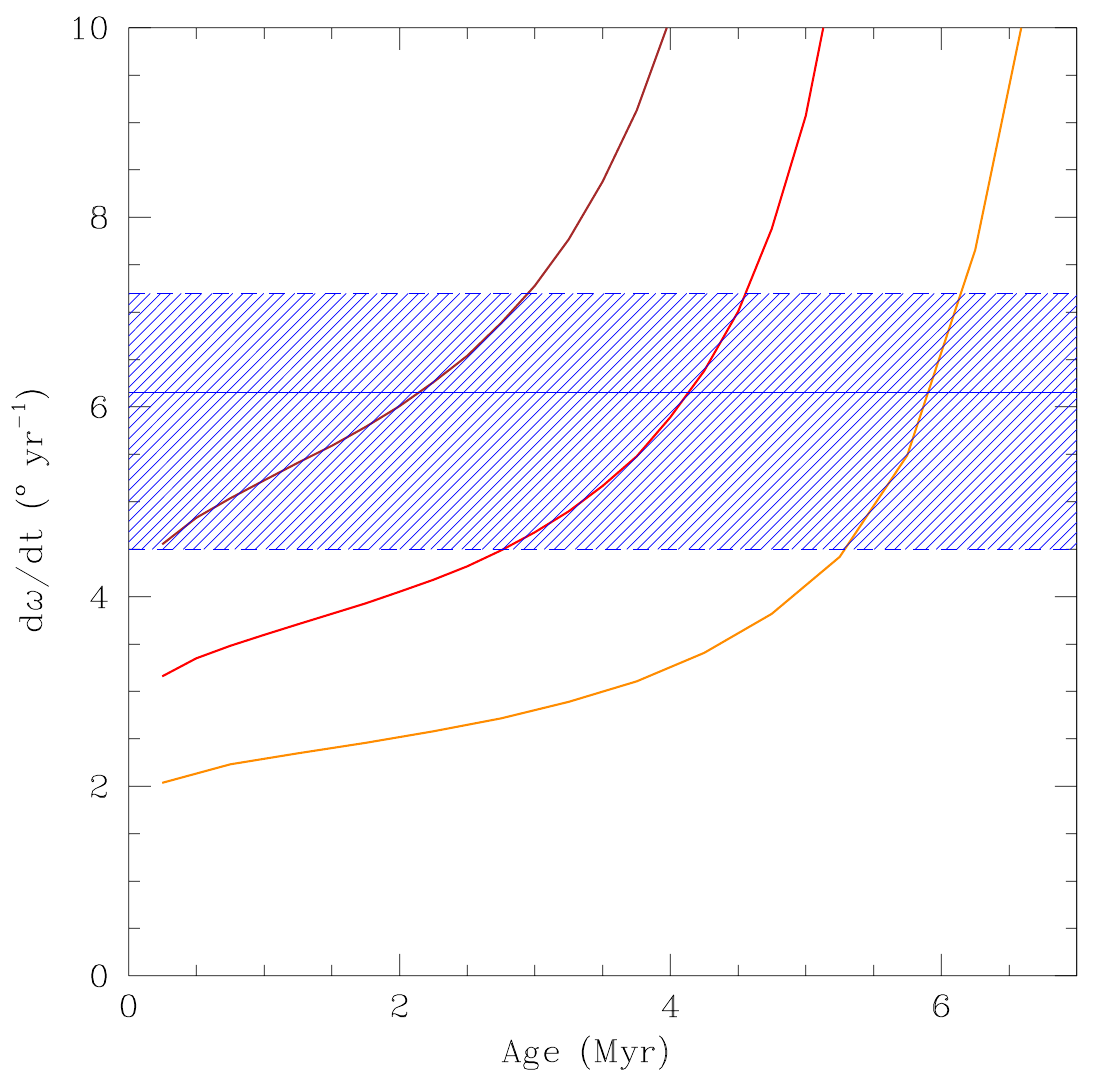}
  \caption{Evolution of the total $\dot{\omega}$ (Newtonian plus general relativity) as a function of age. The red curve displays this evolution for solar-metallicity stars of 26.2\,M$_{\odot}$ and 20.4\,M$_{\odot}$ in a binary with $e = 0.15$ and $P_{\rm anom} = 3.95984$\,d according to the models of \citet{Cla19}. The solid blue line shows our best estimate of $\dot{\omega}$ for  BD+60$^{\circ}$\,497, and the hatched area provides the associated uncertainty. The brown and orange curves correspond to systems respectively with stars of masses 30.3\,M$_{\odot}$ + 23.7\,M$_{\odot}$ and 21.9\,M$_{\odot}$ + 17.1\,M$_{\odot}$.\label{age}}
\end{figure}

We can compare the inferred rate of apsidal motion to theoretically expected values. For this purpose, we assume the stellar rotation axes to be aligned with the axis perpendicular to the orbital plane which is seen under a $45^{\circ}$ inclination angle. The contribution of each star to the Newtonian component of $\dot{\omega}$ depends on several parameters: the orbital eccentricity, $\left(\frac{R_*}{a}\right)^5$ that is the ratio of the stellar radius to the semi-major axis at the power five, and the internal structure constants $k_2$ \citep[see Eqs.\,(5) to (7) of][]{RRN16}. Whilst the eccentricity is well constrained, the lack of photometric eclipses prevents us from using well-determined values for the stellar radii. The $\left(\frac{R_*}{a}\right)^5$ factor thus constitutes by far the biggest source of uncertainties in our calculation. Still, to get an order of magnitude estimate, we can rely on radii and internal structure constants from theoretical stellar evolution models. For this purpose, we used the tables of \citet{Cla19} for solar metallicity\footnote{\citet{Sun17} found that the metallicity of IC\,1805 is either solar or mildly subsolar ($[$Fe/H$] = -0.13$).}. If we consider an age of 3.5\,Myr, that is the main-sequence turn-off age of the IC\,1805 cluster determined by \citet{Sun17}, the theoretical radii are $R_p = 8.85$\,R$_{\odot}$ and $R_s = 6.95$\,R$_{\odot}$. We hence estimated a value of $\dot{\omega}_N$ \citep[corrected for the effect of rotation according to][]{Cla24} near $4.90^{\circ}$\,yr$^{-1}$. To this, we need to add the general relativity contribution which can be evaluated from Eq.\,(10) in \citet{RRN16}. This contribution amounts to $0.27^{\circ}$\,yr$^{-1}$. The total theoretical rate of apsidal motion for an age of 3.5\,Myr thus amounts to $5.17^{\circ}$\,yr$^{-1}$.

Alternatively, we can compare the observational value of $\dot{\omega}$ with the theoretically expected evolution of this quantity as a function of age. For this purpose, we again used the model predictions of \citet{Cla19} along with the correction to $k_2$ for the effects of rotation following the formalism of \citet{Cla24}. For the latter correction, we assumed that the rotational and orbital axes are aligned and are both seen under an inclination of $45^{\circ}$. The results are shown by the red line in Fig.\,\ref{age}. Comparing the theoretical values of $\dot{\omega}$ with our determination results in an age estimate of $4.13^{+0.42}_{-1.37}$\,Myr. \citet{Sun17} evaluated an age spread of $\leq 1.5$\,Myr about the main-sequence turn-off age of 3.5\,Myr for the massive star population in IC\,1805. Our evaluation is thus in reasonable agreement with the cluster age.

Propagating the uncertainties on $m_p\,\sin^3{i}$ and $i$ yields $m_p = 26.2^{+4.1}_{-4.4}$\,M$_{\odot}$. We repeated the above procedure adopting the extreme values of $m_p$ and assuming an $m_s$ consistent with the observational $q$ value. The results are shown by the orange and brown curves in Fig.\,\ref{age}. The more massive the stars, the earlier the system reaches the observational value of $\dot{\omega}$. For the most extreme masses considered here, ages between 0.25\,Myr and 6.25\,Myr would agree within the uncertainties with our observational estimate of $\dot{\omega}$.  

\subsection{CoMBiSpeC simulations \label{sect:combispec}}
An important question is whether the apparent changes in the spectral properties between the 2002-2003 and 2018-2022 epochs reflect another subtle manifestation of the apsidal motion. According to our best orbital solution, the longitude of periastron changed from around $145^{\circ}$ in the earlier epoch to about $250^{\circ}$ in the later epoch. Hence, in 2002-2003, periastron occured at times when the secondary's hemisphere that faces the primary (and thus undergoes the strongest heating and reflection effects) was turned away from us. This constrasts with the more recent epoch where this hemisphere was facing the observer at periastron passage. Hence apsidal motion adds to the aspect changes of the system between the two epochs. 

\begin{figure}
    \includegraphics[angle=0, width=8.5cm]{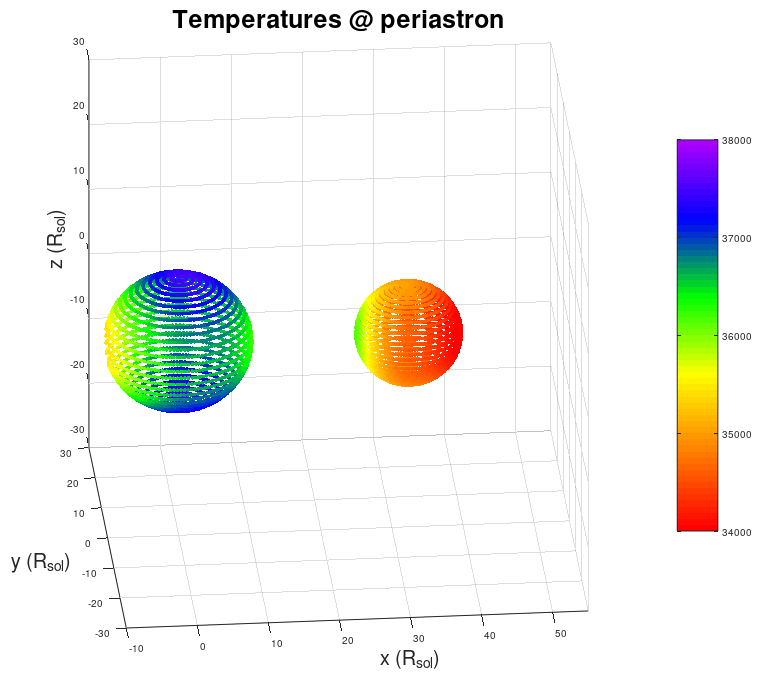}
    \includegraphics[angle=0, width=8.5cm]{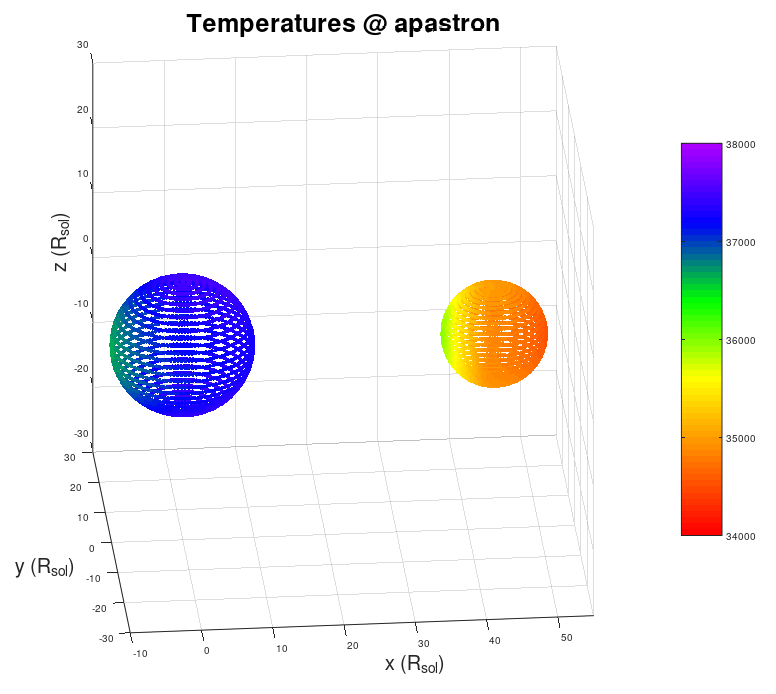}
  \caption{Surface temperature distribution for the stars of the BD+60$^{\circ}$\,497 binary system as computed with CoMBiSpeC at periastron (top) and apastron (bottom). The primary and secondary input radii were taken to be respectively 9.8\,R$_{\odot}$ and 7.3\,R$_{\odot}$, and their effective temperatures are respectively 37.5\,kK and 35\,kK.\label{mutualheating}}
\end{figure}

To quantify the mutual heating effects, we used the CoMBiSpeC code \citep{Pal12,Pal13} to simulate the temperature distribution at the surface of both stars at different orbital phases. CoMBiSpeC discretises the surface of each star into a finite grid of elements. For each cell, the local acceleration of gravity and surface temperature are computed based on the binary gravitational potential accounting for radiation pressure, gravity darkening, and reflection effects \citep[see][for details]{Pal12,Pal13}. For our simulations, we adopted the following stellar parameters: $T_{\rm eff, p} = 37.5$\,kK, $T_{\rm eff, s} = 35$\,kK, $M_{p} = 26.2$\,M$_{\odot}$, $M_{s} = 20.4$\,M$_{\odot}$, $R_p = 9.8$\,R$_{\odot}$ and $R_s = 7.3$\,R$_{\odot}$. The effective temperatures are taken from \citet{Mar05} accounting for our spectral classification, and the radii are inferred from the models of \citet{Cla19} for an age of 4.13\,Myr that best matches the observationally determined rate of apsidal motion. The synthetic surface temperature distributions at periastron and apastron are shown in Fig.\,\ref{mutualheating}.

Owing to gravity darkening and mutual heating, both stars display a non-uniform temperature distribution at their surface. For the secondary the difference in temperature between the hottest and the coolest areas reaches values between 1.4\,kK at apastron and 2.4\,kK at periastron in our simulation. In comparison, for the primary these differences range between 0.8\,kK and 2.1\,kK. The temperatures along the primary's equator reach their lowest values at periastron because, at this orbital phase, the star fills up a larger fraction of its Roche lobe. This results in a lower effective surface gravity\footnote{In a close binary system, the surface gravity is given by the gradient of the binary gravitational potential \citep{Pal12,Pal13}.} and thus a lower effective temperature in the equatorial region. Over the hemisphere that faces the secondary, this effect is only partially compensated by the stronger irradiation from the secondary star at periastron. At this orbital phase, the hemispheres of both stars that face each other have nearly identical temperatures.

Because of the irradiation by the hotter primary, the hottest side of the secondary is always the hemisphere facing the primary. The CoMBiSpeC simulations predict a maximum temperature on this hemisphere that remains nearly constant as a function of orbital phase (near 36.1\,kK). The difference between periastron and apastron only amounts to 200\,K. At first sight, this result might seem counterintuitive as one would expect enhanced heating at periastron. However, as outlined above, the enhanced cooling effect of gravity darkening around periastron, lowers the relevant primary temperature. The lower primary temperature compensates the reduction in separation between the stars that would otherwise boost the mutual heating effect. This also explains why the lowest secondary surface temperatures are predicted at periastron: as a result of gravity darkening, the rear side of the secondary has its temperature varying between 33.8\,kK at periastron and 34.5\,kK at apastron. 

We also computed the surface temperature distributions for the orbital configurations corresponding to the spectra illustrated in Fig.\,\ref{montage}. The results are illustrated in Fig.\,\ref{mutual}. The quadrature phases in 2002-2003 correspond to configurations where the lower limit of the secondary's temperature distribution is lower by 300 to 500\,K than at the quadrature phases of the 2018-2022 epoch. This is due to the fact that the quadrature phases in 2002-2003 happened at smaller orbital separations. Whilst this effect goes into the right direction, it seems questionable whether such a modest difference in temperature could account for the differences in the properties of the disentangled spectra. The simulations predict also a $\sim 500$\,K lower apparent temperature for the primary star at quadrature phases in 2002-2003 compared to the same phases in 2018-2022.  

\begin{figure}
  \includegraphics[angle=0, width=8.5cm]{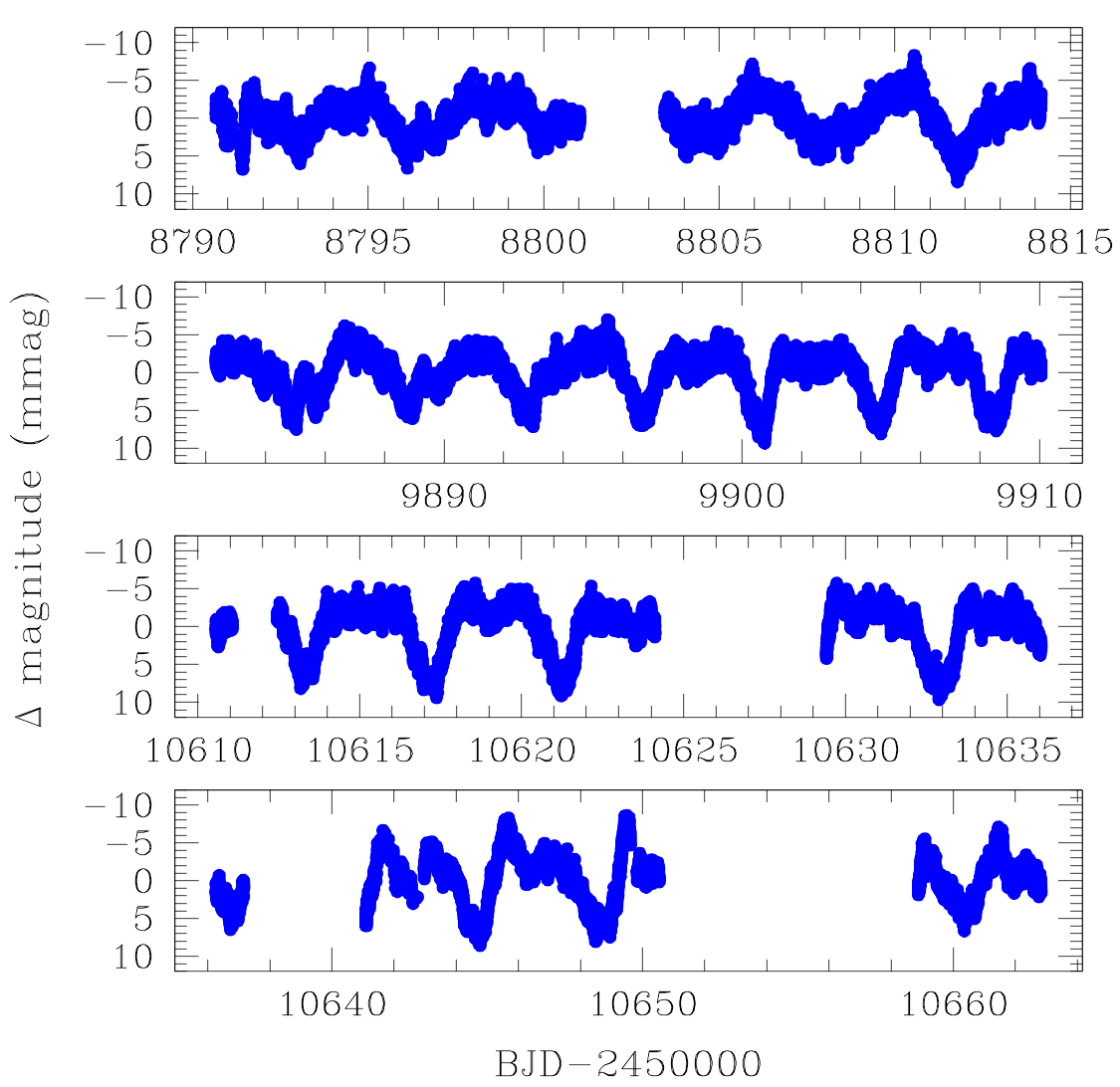}
  \caption{{\it TESS} light curves of BD+60$^{\circ}$\,497. The panels illustrate, from top to bottom, the PDC photometry from Sectors 18, 58, 85, and 86. \label{TESSlc}}
\vspace*{3mm}
  \includegraphics[angle=0, width=8.5cm]{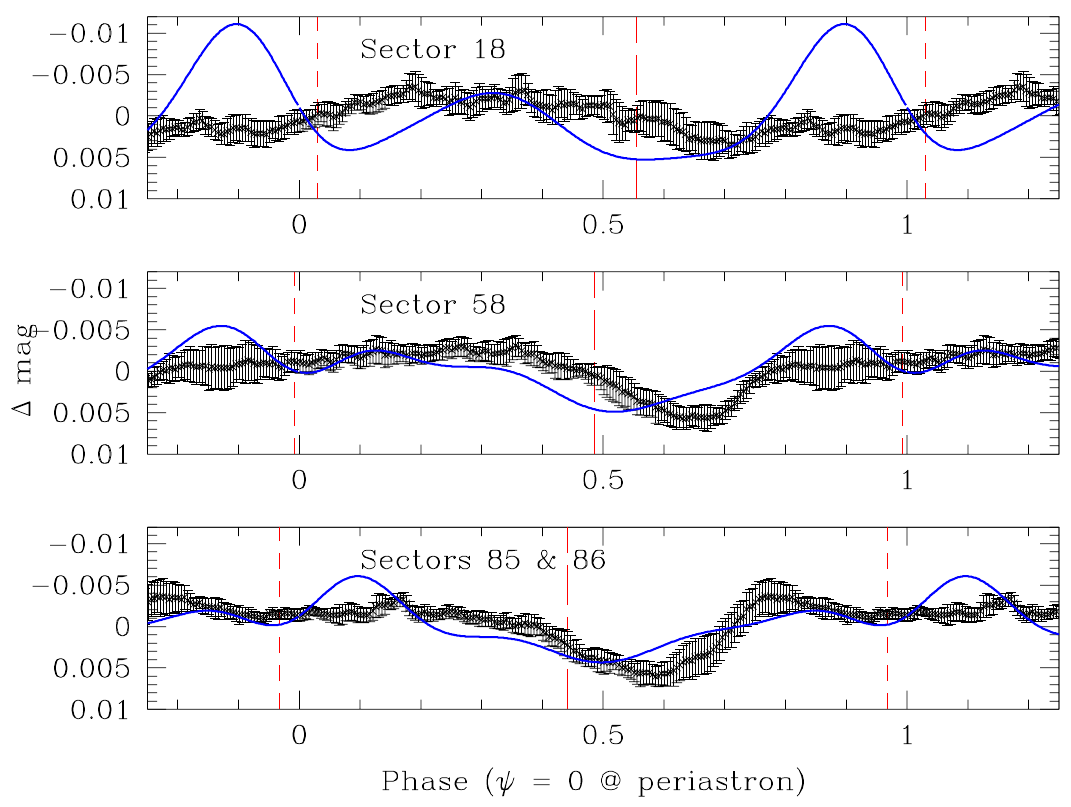}
  \caption{Phase-folded {\it TESS} light curves of BD+60$^{\circ}$\,497 along with our best eBEER + TEO model (blue curve, see Sect.\,\ref{sectTEO}). Phase 0.0 corresponds to periastron passage ($t_0 = 2456682.7$, $P_{\rm anom} = 3.95984$\,d, see Table\,\ref{omegadotfit}). Conjunction with the primary star in front occurred at phases $0.030$, $0.992$ and $0.967$ respectively in Sectors 18, 58, and 85 and 86 (red short-dashed vertical lines). Conjunction with the secondary star in front occurred at phases $0.554$, $0.486$ and $0.441$ respectively in Sectors 18, 58, and 85 and 86 (red long-dashed vertical lines). The eBEER model was computed using the RV dataset 2 orbital solution from Table\,\ref{omegadotfit}. We assumed an orbital inclination $i = 45^{\circ}$ and radii of 9.8 and 7.3\,R$_{\odot}$ for the primary and secondary stars. The albedo coefficient of both stars was 0.3. The best-fit TEOs have $A_2 = 4.4$\,mmag, $\psi_2 = 2.39$\,rad, $A_3 = 2.4$\,mmag and $\psi_3 = 3.39$\,rad. \label{TEOfit}}
\end{figure}

\section{Photometry \label{Photom}}
As expected from the low orbital inclination (near $45^{\circ}$) inferred from our RV analysis, the {\it TESS} light curves (Fig.\,\ref{TESSlc}) do not exhibit any strong photometric eclipses that could help constrain the rate of apsidal motion. Instead, the light curves display only a low-level modulation with a single shallow minimum occurring once per orbital cycle. Between Sectors 18 and 86, there is a progressive change in the shape of the light curve. To highlight those differences, we built a phase-folded light curve for each {\it TESS} sector. We binned the data into 200 intervals spanning 0.005 in orbital phase. Phase 0.0 was taken as the time of periastron according to the ephemerides derived from our RV study accounting for apsidal motion (RV dataset 2, Table\,\ref{omegadotfit}). For each phase bin, the error on the average magnitude was evaluated as
$\sqrt{\sigma^2_{\rm disp} + \sigma^2_{\rm err}}$,
where $\sigma_{\rm disp}$ stands for the dispersion of the individual data points falling into that phase bin and $\sigma_{\rm err}$ is the mean photometric error of the data points in that phase bin. The corresponding light curves are illustrated in Fig.\,\ref{TEOfit}.

Although there exists some scatter in the phase-folded data points, the light curves from the different sectors are not identical as already noted above. In the Sector 18 data, the minimum is not very well defined and consists in a broad feature made of two consecutive minima of unequal depths. During Sector 58, a single minimum is observed, which appears better defined (sharper and deeper) although it clearly remains asymmetric with a slow brightness decrease followed by a faster recovery. Finally, the combined light curve of Sectors 85 and 86 again yields a sharper and deeper minimum followed immediately by a sharp maximum, which is more prominent in Sector 86 but hardly visible in Sector 85. 

The folded light curves indicate that the shallow minima do not correspond to grazing eclipses which would be expected to occur at the conjunction phase closest to periastron passage. At the epochs of the {\it TESS} observations, $\omega$ was in the range between $255^{\circ}$ (Sector 18), $274^{\circ}$ (Sector 58), and $286^{\circ}$ (Sector 86). Hence, the conjunction phases closest to periastron were expected around phases $0.030$ (Sector 18), $0.992$ (Sector 58) and $0.967$ (Sectors 85 and 86). Instead, Fig.\,\ref{TEOfit} shows the shallow minima to occur around phase 0.6 -- 0.7. 

\begin{figure}
  \includegraphics[angle=0, width=8.5cm]{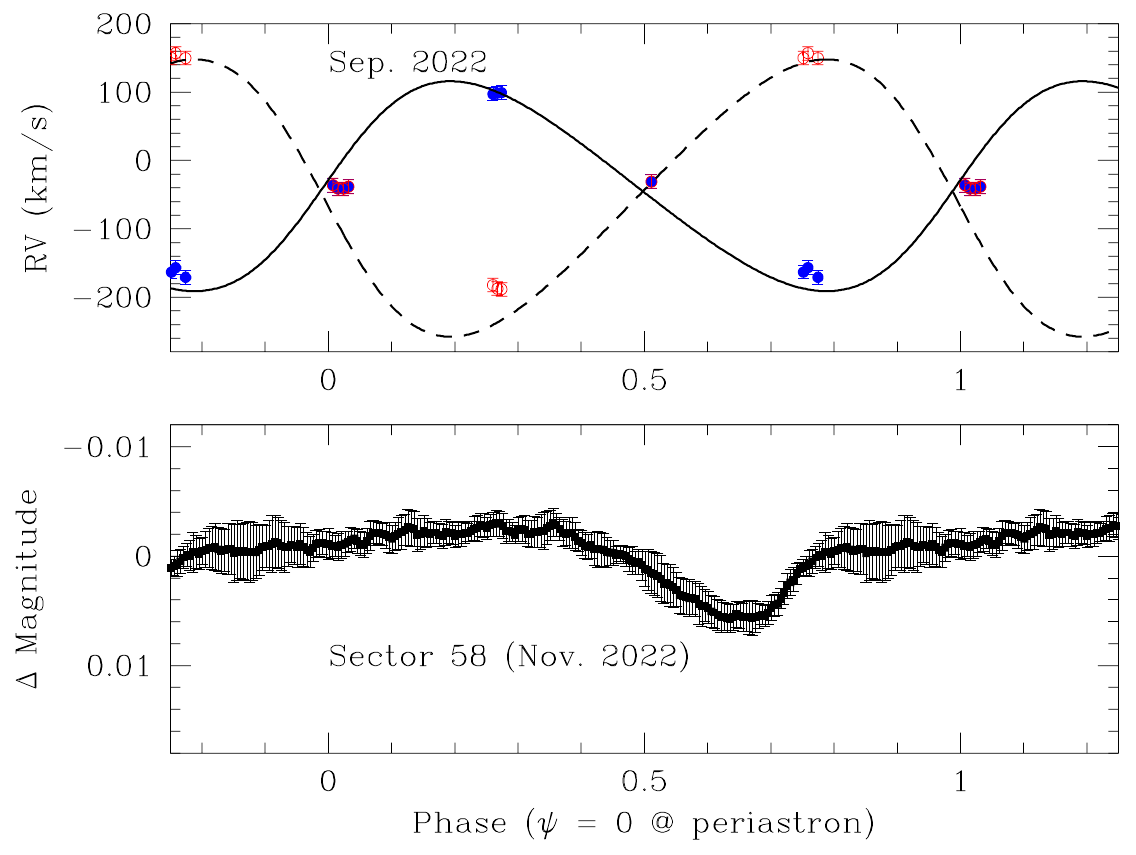}
  \caption{Orbital solution and light curve of BD+60$^{\circ}$\,497 in the autumn 2022. Top panel: RV data along with our best-fit orbital solution accounting for apsidal motion (see Table\,\ref{omegadotfit}). Primary RVs are shown in blue, secondary RVs in red. Only the primary RVs at phases with a RV separation larger than 200 km\,s$^{-1}$ were included in the RV fit. RVs at phases when the lines are heavily blended are shown here but were not used in the fit. Bottom panel: {\it TESS} Sector 58 light curve folded with the same ephemerides. Phase 0.0 is periastron. Conjunctions occurred at phases 0.992 (primary in front) and 0.486 (secondary in front).\label{compa}}
\end{figure}

To check that there is no significant phase shift between the RV and photometric data we considered the RV data collected in August 2019 and September 2022. These data were taken within about 2 months of the photometric observations of Sectors 18 and 58. Over such a time interval, the uncertainties on $P_{\rm anom}$ would result in a phase shift of at most 0.001. Figure\,\ref{compa} illustrates the RV data and our best-fit orbital solution accounting for apsidal motion along with the {\it TESS} curve from Sector 58 folded with the same ephemerides. This figure clearly shows that the minimum in the light curve is offset from conjuction phases and occurs at a time when the primary moves towards the observer and the secondary moves away from the observer.

\begin{figure*}
  \begin{minipage}{5.7cm}
    \includegraphics[angle=0, width=5.7cm]{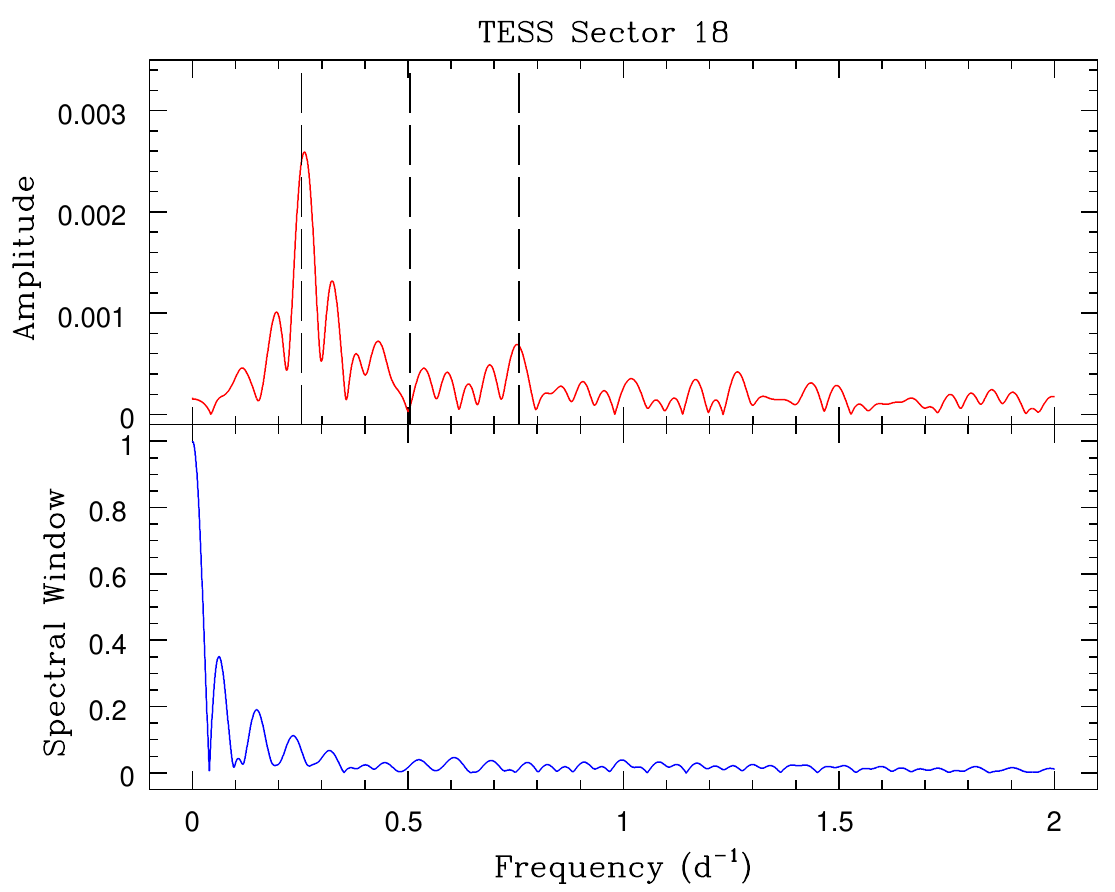}
  \end{minipage}
  \hfill
  \begin{minipage}{5.7cm}
    \includegraphics[angle=0, width=5.7cm]{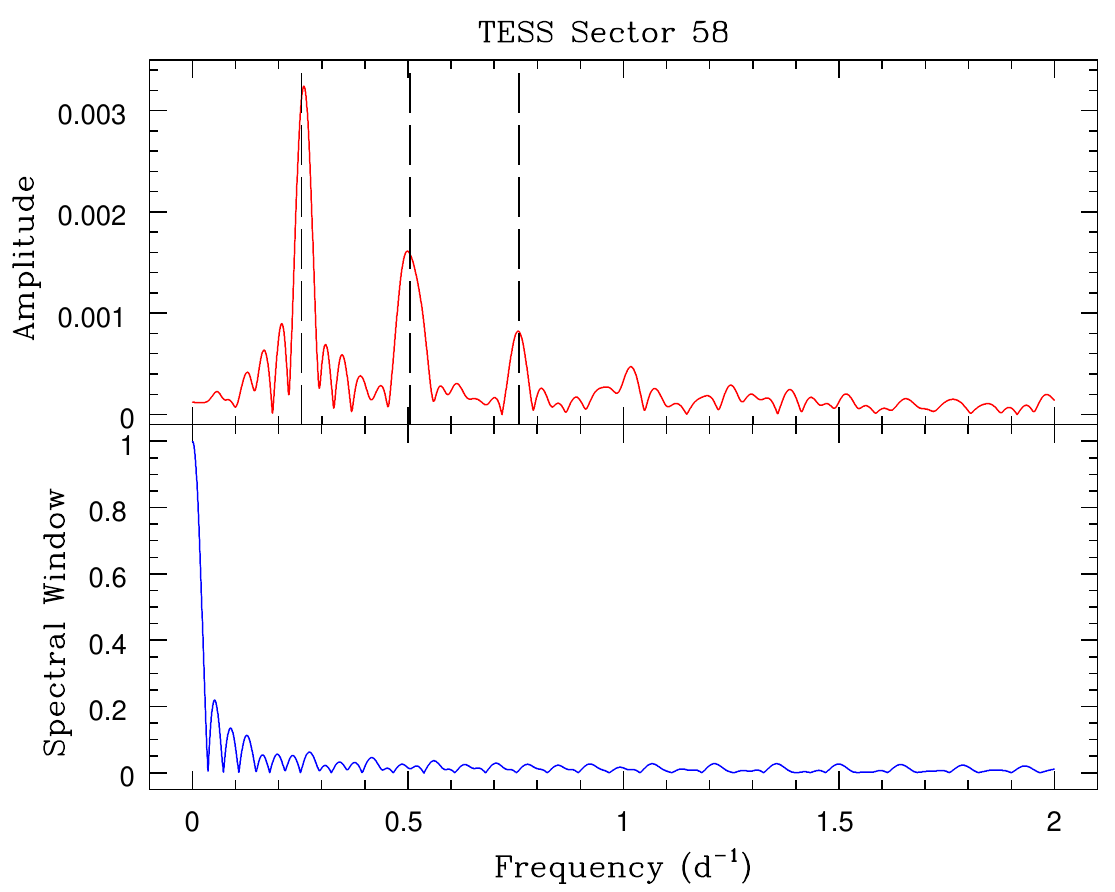}
  \end{minipage}
  \hfill
  \begin{minipage}{5.7cm}
    \includegraphics[angle=0, width=5.7cm]{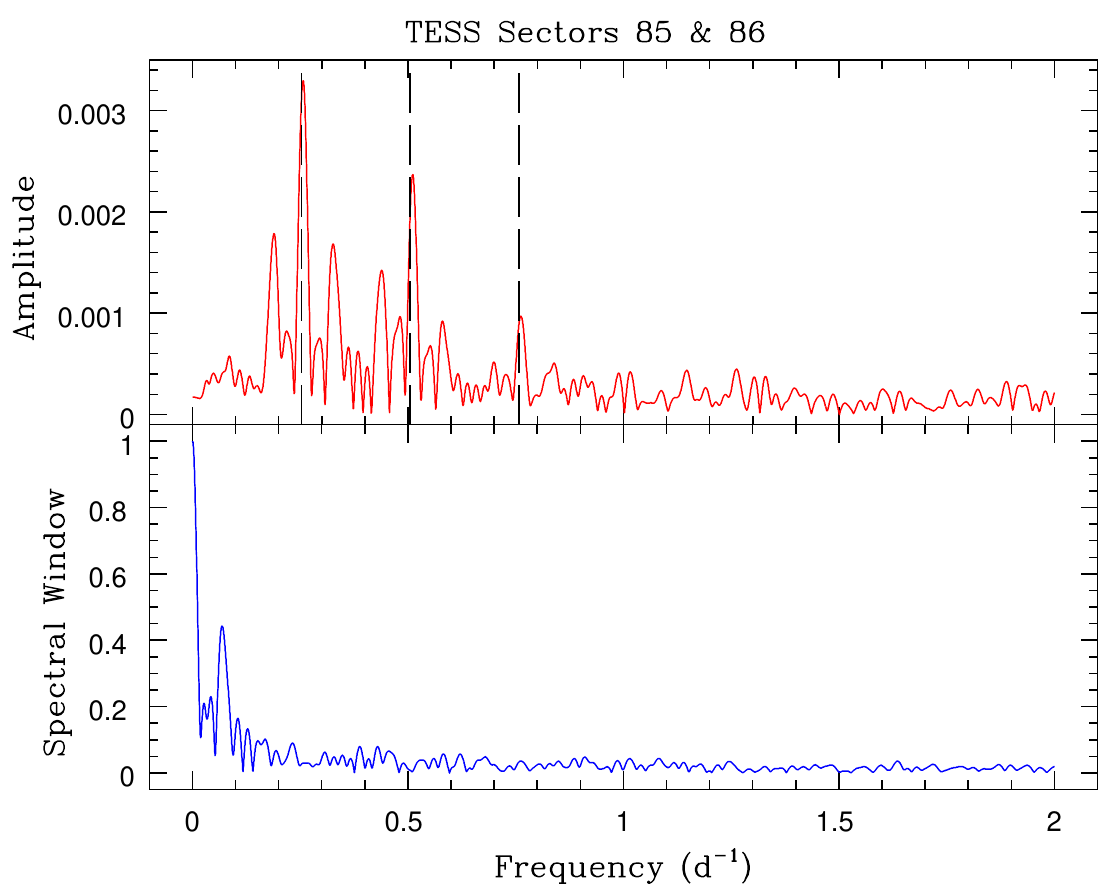}
  \end{minipage}
  \caption{Fourier amplitude spectra of the {\it TESS} light curves of BD+60$^{\circ}$\,497. For each sector or combination of sectors, the upper panel shows the Fourier spectrum of the photometric time series, whereas the lower panel displays the spectral window function associated with the sampling. The dashed vertical lines indicate the orbital frequency corresponding to $P_{\rm anom}$ and its $n=2$ and $n=3$ harmonics.\label{spTESS}}
\end{figure*}

\begin{table*}
  \caption{Results of Fourier analyses of {\it TESS} light curves. \label{FourTess}}
  \begin{tabular}{c c c c c c c c c}
    \hline
    Sector & $\Delta\,\nu_{\rm nat}$ & \multicolumn{2}{c}{Dominant peak} & \multicolumn{4}{c}{Harmonics} & Fundamental frequency \\
    & (d$^{-1}$) & $\nu_1$ (d$^{-1}$) & Amp$_1$ (mmag) & $\nu_2$ (d$^{-1}$) & Amp$_2$ (mmag) &  $\nu_3$ (d$^{-1}$) & Amp$_3$ (mmag) & (d$^{-1}$) \\
    \hline
    18 & 0.043 & 0.2596 & 2.59 & -- & -- & 0.7534 & 0.69 & $0.2578 \pm 0.0043$ \\
    58 & 0.036 & 0.2583 & 3.24 & 0.4967 & 1.61 & 0.7556 & 0.83 & $0.2570 \pm 0.0036$ \\
    85 \& 86 & 0.019 & 0.2566 & 3.30 & 0.5107 & 2.37 & 0.7618 & 0.97 & $0.2558 \pm 0.0019$ \\
    \hline
  \end{tabular}
  \tablefoot{The column with $\Delta\,\nu_{\rm nat}$ lists the natural width of the peaks. The last column presents the fundamental frequency determined as the amplitude-weighted mean of $\nu_1$ and of $1/n$ of the $n^{\rm th}$ harmonic frequency. The second harmonic, $n = 2$, was only included for Sects.\ 85 and 86.}
\end{table*}

To further quantify the behaviour of the light curves, we performed a Fourier analysis. We determined the frequency content of the data of each individual sector using the discrete Fourier power spectrum method developped by \cite{HMM} and amended by \cite{Gos01}. This algorithm explicitly accounts for the irregularities (gaps) in the sampling of the data. The Fourier spectrum, computed up to the Nyquist frequency of 360\,d$^{-1}$, is essentially empty above 2\,d$^{-1}$. Hereafter, we thus focus on its low frequency content. Whilst there is clearly some low-level intrinsic variability, the strongest signals are related to the orbital cycle.

The results are illustrated in Fig.\,\ref{spTESS} and summarised in Table\,\ref{FourTess}. The dominant peak near 0.26\,d$^{-1}$ is seen in all three Fourier spectra. The visibility of its harmonics changes with epoch. At first, in Sector 18, only the third harmonic frequency is detected. In Sector 58, the second harmonics also shows up, but the associated peak appears significantly broader than expected (full width at half maximum nearly twice the natural width), suggesting that it might correspond to a blend with another frequency. Finally, the Fourier analysis of the combined dataset from Sectors 85 and 86 reveals second and third harmonics with strongly increased amplitudes compared to previous sectors. The width of the second harmonics is now consistent with the natural width. The last column of Table\,\ref{FourTess} shows the amplitude-weighted mean of $\nu_1$ and of $1/n$ of the $n^{\rm th}$ harmonic frequency. For Sectors 18 and 58, the contribution of the harmonics to this mean was restricted to $n = 3$, whilst $n = 2$ and $n = 3$ were considered for the combined dataset from Sectors 85 and 86. The uncertainty on the resulting fundamental frequency is taken to be $\Delta\,\nu_{\rm nat}/10$. 

Formally speaking, the fundamental frequencies found for each sector are close to - but different from - the anomalistic orbital frequency (0.2525\,d$^{-1}$) or the sidereal orbital frequency (0.2526\,d$^{-1}$). To check whether this could be due to genuine period variations, we performed dedicated simulations with the \texttt{binary} package of the Modules for Experiments in Stellar Astrophysics \citep[MESA;][]{Pax11, Pax13, Pax15, Pax18, Pax19, Jer23}, which is a one-dimensional stellar structure and evolution code. Using this tool, we verified if the interplay between tides and wind mass loss of both components may result in significant changes in the orbital period of BD+60$^\circ$\,497. The details of these computations can be found in Appendix\,\ref{app:mesa-input}. Depending on the individual initial rotation rates of the components, the system may experience either positive or negative $\dot{P}_{\rm orb}$. However, these experiments gave $\frac{|\dot{P}_{\rm orb}|}{P_{\rm orb}}$ of order $10^{-8}$ to $10^{-7}$\,yr$^{-1}$. These theoretical values thus imply $|\dot{P}_{\rm orb}| \leq 4 \times 10^{-7}$\,d\,yr$^{-1}$, several orders too small to explain the differences between fundamental photometric frequencies and orbital frequency as a result of genuine orbital period changes. 

\section{Discussion}
There are a number of effects that can affect the photometry and spectroscopy of a system such as BD+60$^{\circ}$\,497. In a non-eclipsing eccentric binary, photometric modulations can occur because of ellipsoidal variations, mutual reflection effects and Doppler beaming \citep[e.g.][and references therein]{Eng20}. In addition, TEOs with frequencies close to resonance with harmonics of the orbital frequency can also play a non-negligible role in the photometric variability \citep{Kol23}. Because the binary components are O-type stars, they should have strong stellar winds that interact in between the stars. This wind-wind collision can impact the light curve \citep[e.g.][]{Ant24,Kol24}. Additional effects could arise in a hierarchical triple system or from asynchronous rotation of a spotted rotator. Hereafter, we discuss the pros and cons of these various scenarios. 

\subsection{Proximity effects}
\begin{figure}
  \includegraphics[angle=0, width=8.5cm]{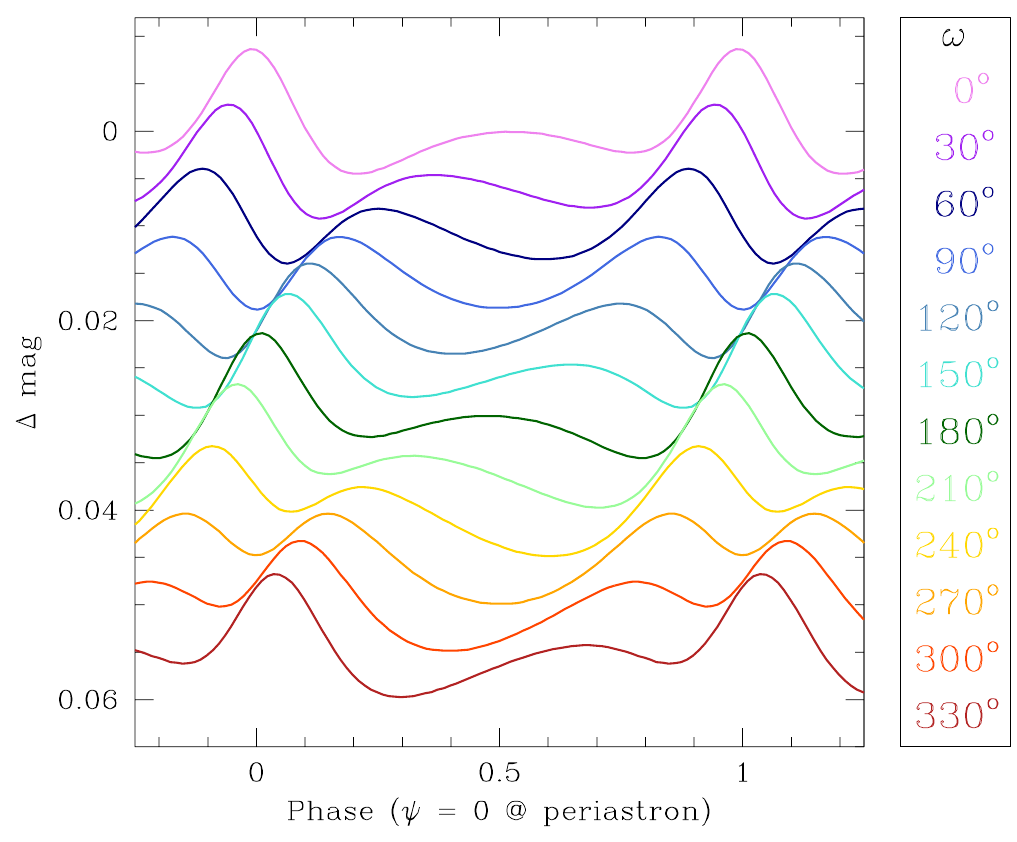}
  \caption{Variation in the theoretical light curve computed with the {\it Nightfall} code as a function of $\omega$. The parameters of the model were taken as $i = 45^{\circ}$, $fill_p = fill_s = 0.6$, $T_{{\rm eff}, p} = 37.5$\,kK, $T_{{\rm eff}, s} = 32.5$\,kK, $e = 0.15$.\label{montageLC}}
\end{figure}

Binary codes such as {\it Nightfall} \citep{Wichmann} and {\it PHOEBE} \citep{Prs18}, or the semi-analytical eccentric BEaming Ellipsoidal and Reflection formalism \citep[eBEER,][]{Eng20}, all predict a double-wave light curve for an eccentric non-eclipsing binary. This shape is due to the ellipsoidal deformations that result in a minimum at each conjunction phase, whilst beaming and reflection result in a single wave modulation \citep[e.g.][]{Naz23}. The exact morphology of the synthetic light curve depends on the relative importance of reflection and ellipsoidal effects, which in turn depends on the assumed reflection coefficient and the adopted limb-darkening and gravity darkening coefficients \citep{Eng20}. Among the {\it TESS} light curves of BD+60$^{\circ}$\,497 (Fig.\,\ref{TEOfit}), the data from Sectors 85 and 86 come closest to this description. Allowing for phase shifts and letting the codes adjust the secondary temperature, the filling factors of both stars and the value of $\omega$, we found indeed some {\it Nightfall} and {\it PHOEBE} models that could reasonably reproduce the shape of the Sectors 85 and 86 light curve (see Fig.\,\ref{appendLC}). This model predicts two photometric minima at conjunction phases: the deepest and broadest one after apastron and a narrower and shallower minimum near periastron. However, there are several major caveats here. First of all, matching the amplitude of the observed minimum requires significantly smaller stars than expected from their spectral type. Setting instead the radii to their expected values results in a minimum twice as deep as observed. Second, matching the phase of the deepest minimum either requires a shift in phase of $-0.12$, which is at odds with the RV variations as shown in Sect.\,\ref{Photom}, or a value of $\omega$ around $255^{\circ}$ instead of $286^{\circ}$, expected at that epoch from our apsidal motion study. Figure\,\ref{montageLC} illustrates the expected behaviour of the light curve as $\omega$ evolves from 0 to $330^{\circ}$. The phase of the deepest minimum and the relative height of the two maxima of the light curve vary with $\omega$. The maxima have equal strengths for $\omega = 90^{\circ}$ and $\omega = 270^{\circ}$. At $\omega = 255^{\circ}$, the phase of the predicted minimum better matches the observations and the strongest maximum is expected at phase $0.9$ (see Fig.\,\ref{appendLC}). However, such a large error on the longitude of periastron would require pushing the uncertainties on $\omega_0$ and $\dot{\omega}$ to their most extreme values. What is even worse is that the observed epoch-dependence of the light curve is at odds with the dependence on $\omega$ shown in Fig.\,\ref{montageLC}, as can be seen on Fig.\,\ref{appendLC}. In Sectors 18 (resp.\ 58), the expected value of $\omega$ would be about $23^{\circ}$ (resp.\ $9^{\circ}$) lower than in Sectors 85 and 86. For such values of the longitude of periastron, we expect a strong peak near phase 0.90 -- 0.92, which clearly is not observed.

Therefore, proximity effects alone cannot explain the shape of the observed photometric light curve. Given the properties of the system, such effects ought to exist with an amplitude larger than that of the observed variations. Hence, an important question is what could actually inhibit such effects or hide them. In the next subsections, we consider various possibilities.

\subsection{TEOs \label{sectTEO}}
The picture could be blurred if BD+60$^{\circ}$\,497 exhibits TEOs in addition to ellipsoidal variations, reflections and beaming. Our Fourier analysis of the light curves did not unveil harmonics beyond $n = 2$ and $n = 3$, except possibly for Sectors 85 and 86, where the $n = 4$ harmonic is likely present at 1.0168\,d$^{-1}$ (amplitude 0.0011\,mag), although it stands out only very weakly against the noise level. The $n = 2$ and $n = 3$ harmonics fall into a frequency regime where $l = 2$ stellar eigenmodes should have a significant overlap integral with the forcing frequency. Hence, they could indeed excite significant TEOs that would come on top of the proximity effects.   

To test this scenario, we simulated a grid of synthetic light curves including the proximity effects computed with the eBEER \citep{Eng20} model as well as TEOs with $n=2$ and $n=3$. For the proximity effects we assumed the $\omega$ values computed from our orbital solution including apsidal motion. The only free parameter associated with the eBEER model was the reflection coefficient taken equal for both stars. Each TEO brings in another two free parameters which are its amplitude $A_n$ and phase constant $\psi_n$. We then attempted to adjust the light curves from all three epochs with a single set of amplitudes and phase constants using an expression of the kind
\begin{eqnarray}
  m_{\it TESS}(t) & = & {\rm eBEER}(\omega(t),\psi(t)) + A_2\,\cos{(4\,\pi\,\psi(t) + \psi_2)} \nonumber \\
  & & + A_3\,\cos{(6\,\pi\,\psi(t) + \psi_3)},
\end{eqnarray}
where $\psi(t)$ is the orbital phase measured from periastron. We varied the albedo coefficient from 0.1 to 1.0 in steps of 0.1. This parameter appears to be anti-correlated with the amplitude $A_3$ of the $n = 3$ TEO: the lower the albedo coefficient, the higher $A_3$. The same rule applies to $A_2$ although here the relationship is less well defined. The smallest residuals were obtained for a surprisingly low albedo coefficient of 0.3. The corresponding curves are shown in Fig.\,\ref{TEOfit}. The agreements are clearly not perfect, but provide reasonable fits for Sector 58 and Sectors 85 and 86. However, for Sector 18 the model predicts variations between phases 0.95 and 1.35 that clearly are not observed.

We thus conclude that TEOs with $n = 2$ and $n = 3$ could alleviate some of the difficulties encountered when fitting the light curves. However, fitting all three epochs of {\it TESS} observations with the same set of TEO parameters remains challenging.  

\subsection{Stellar wind effects}
The presence of stellar winds in the BD+60$^{\circ}$\,497 system was diagnosed thanks to low-dispersion {\it IUE} spectra that revealed prominent C\,{\sc iv} $\lambda\lambda$ 1548, 1551 P-Cygni profiles. \citet{Bur79} estimated a terminal wind velocity $v_{\infty} = 3600$\,km\,s$^{-1}$, whilst \citet{Llo82} evaluated a mass-loss rate of $\dot{M} = 6.5 \times 10^{-7}$\,M$_{\odot}$\,yr$^{-1}$. These early estimates neither accounted for the binarity of the system nor for clumps. Given their spectral types, each component of the system is likely to have a stellar wind, but we expect the wind of the hotter and more luminous primary to be the denser and more energetic one. Using the \citet{Bjo23} recipe we estimated $\dot{M}_p = 1.2 \times 10^{-7}$\,M$_{\odot}$\,yr$^{-1}$ for the primary, and $\dot{M}_s = 5.6 \times 10^{-9}$ -- $1.9 \times 10^{-8}$\,M$_{\odot}$\,yr$^{-1}$ for the secondary depending on its effective temperature. \citet{Bjo21} found a rather wide range of values between 2.5 and 5.5 for the ratio $v_{\infty}/v_{\rm esc}$ where $v_{\rm esc}$ is the effective escape velocity. If we adopt a value of 4 for this ratio, we obtain asymptotic wind velocities of 3500\,km\,s$^{-1}$ for the primary and 3800\,km\,s$^{-1}$ for the secondary. These values are in good agreement with the observational result of \cite{Bur79}.    

The presence of stellar winds could impact the photometric light curves as a result of atmospheric eclipses where free electron scattering removes some part of the photons coming from the star that is located behind the wind of its companion. This effect is expected to be strongest at conjunction when the star with the stronger wind is in front. Therefore, given the orbital configuration of BD+60$^{\circ}$\,497 at the time of the {\it TESS} observations, atmospheric eclipses should be maximum near periastron. However, the minima of the light curves are not observed around those phases. Moreover, if we evaluate the typical optical depth due to electron scattering $\tau_0 = \sigma_T\,\frac{\dot{M}}{4\,\pi\,m_p\,\mu_e\,a\,v_{\infty}}$ \citep[Eqs.\ (5) and (6) of][]{Ant24} for the primary wind, it only amounts to 0.0003, which is clearly not sufficient to explain the observed variations. Hence, atmospheric eclipses are unlikely to play a significant role here.

Another consequence of the winds is the existence of a wind-wind interaction zone. Indeed, the stellar winds of both stars collide in between the stars. As the kinetic energy of the winds is converted into heat at the shock, this can lead to a strong phase-dependent X-ray emission on top of the intrinsic emission of each star \citep[see e.g.][for a review]{RauNaz}. In our case, there is no evidence of such an X-ray overluminosity: BD+60$^{\circ}$\,497 has $\log{\left(\frac{L_{\rm X}}{L_{\rm bol}}\right)} = -7.03 \pm 0.06$ following nearly exactly the canonical relation for the intrinsic X-ray emission of O-type stars \citep{Rau16}. However, the lack of strong X-ray emission does not rule out the existence of a wind-wind interaction. Indeed, given the short orbital separation, the stellar winds will not have the time to accelerate to $v_{\infty}$ before they collide. The wind-wind interaction zone should be in the radiatively cooling regime emitting mostly at longer wavelengths (EUV, UV). From our estimates of the mass-loss rates, we can estimate a wind momentum ratio $\eta = \frac{\dot{M_p}\,v_{\infty, p}}{\dot{M_s}\,v_{\infty, s}} \simeq$ 5.5 -- 20. Hence, the primary wind clearly dominates the wind of the secondary, and the collision between the two must be located very close to the surface of the secondary star. This could lead to a back-warming of the hemisphere of the secondary facing the apex of the wind-wind collision zone. Because of the Coriolis deflection, the shock region, and thus the part of the surface affected by the back-warming effect would be skewed by an angle, $\theta$, with respect to the binary axis. Following \citet{Par08}, we have $\tan{\theta} = \frac{v_{\rm orb}}{v_{\infty}}$, where $v_{\rm orb}$ is the orbital velocity of the secondary with respect to the primary and $v_{\infty}$ is the terminal velocity of the slowest wind. From $K_p = 153.5$\,km,s$^{-1}$, $q = 1.28$ and $i \simeq 45^{\circ}$, we can evaluate $v_{\rm orb}$ as 495\,km\,s$^{-1}$. We thus estimated a deflection of about $8^{\circ}$. Such a situation would then lead to a bright spot at the secondary surface which would have its visibility modulated by the orbital motion of the stars. In such a configuration, minimum light would be reached when the bright spot is hidden from our view, that is shortly after conjunction as the secondary starts receding. This is not observed. The skewing angle should shift the minimum away from conjunction, but only by a few hundredths in phase.

A crucial question is whether the power of the wind-wind collision is sufficient to drive a significant heating of the secondary's surface. The power radiated by the wind interaction zone comes from the conversion of the wind's kinetic energy into heat. According to \citet{Pit02}, the power radiated by the dominant wind for the case where $\eta \simeq 10$ is given by $0.033\,\left(\frac{1}{2}\,\dot{M}\,v_{\infty}^2\right)$, and the contribution of the weaker wind is about four times lower. In our case, the total power of the wind interaction zone would amount to about $2 \times 10^{34}$\,erg\,s$^{-1}$, that is 5\,L$_{\odot}$ or $2.6 \times 10^{-5}\,L_{{\rm bol}, p}$. This corresponds to the power emitted in all directions, and thus it provides only an upper limit on the power radiated towards the secondary. This is several orders of magnitude less than the power of order $10^{-2}\,L_{{\rm bol}, p}$ that the secondary receives from the primary via direct illumination. Our CoMBiSpeC simulations indicated that the direct illumination of the secondary by the primary leads to an increase in effective temperature by $\sim 2$\,kK. Hence the small additional contribution due to the wind-wind interaction is unlikely to account for the $\sim 3$\,kK difference in temperature needed to explain the changes in apparent spectral type found in our spectral analysis.

\subsection{Asynchronous rotational modulation}
Assuming the secondary star has a non-uniform surface brightness distribution could help solve both the issue of its epoch-dependent spectroscopic signature and the problem of the shape of the light curves. Because this non-uniform brightness distribution would rotate with the star, the changing aspect of the light curve from one epoch to another could then hint at a slightly asynchronous rotation. Bright surface spots with temperature differences of a few kiloKelvin could arise from the surface emergence of magnetic fields generated in a subsurface iron convection zone \citep{Can11}. Still, to explain the changing properties of the secondary spectrum in this way, would require spots covering a significant fraction of the stellar surface.

Tidal interactions in binary systems not only tend to circularise the orbit, but also to synchronise stellar rotation and orbital motion \citep[e.g.][]{Hut81}. In an eccentric binary system, such as BD+60$^{\circ}$\,497, the orbital angular velocity varies with orbital phase. Strict synchronisation can thus not be reached until the orbit has circularised. The strongest interactions take place around periastron passage, and one expects the rotational velocities to quickly synchronise with the instantaneous orbital angular velocity at periastron \citep{Hut81}. This corresponds to a period given by $P_{\rm peri-sync} = P_{\rm anom}\,\frac{(1-e)^{3/2}}{\sqrt{1+e}}$. In our case, this period amounts to 2.8937\,d. \citet{Hut81} introduced pseudo-synchronisation as a state where the rotation rate remains constant. The corresponding pseudo-synchronisation period is the result of averaging the tidal effects around periastron and is thus slightly longer than the instantaneous rotation period at periastron \citep[see equation (45) of][]{Hut81}. In our case $P_{\rm pseudo-sync} = 1.205\,P_{\rm peri-sync} = 3.4875$\,d.

From our determinations of the projected rotational velocities, $v\,\sin{i} = 130 \pm 20$\,km\,s$^{-1}$ for the primary and $v\,\sin{i} = 110 \pm 20$\,km\,s$^{-1}$ for the secondary, we can estimate the rotational periods assuming the stellar radii and the inclination of the rotational axes. Regarding the rotational axes, we assume that they are aligned with the axis perpendicular to the orbital plane for which we have estimated $i = 45^{\circ}$. For the radii, we again take $9.8$ and $7.3$\,R$_{\odot}$ respectively for the primary and secondary stars. In this way, we estimate rotational periods of $(2.70 \pm 0.41)$\,d for the primary and $(2.37 \pm 0.43)$\,d for the secondary. These values suggest that BD+60$^{\circ}$\,497 has achieved near synchronisation with the instantaneous rotation period at periastron. The estimated rotational periods of the components correspond to frequencies of $(0.37 \pm 0.05)$\,d$^{-1}$ and $(0.42 \pm 0.07)$\,d$^{-1}$. Neither of these frequencies shows up in our Fourier periodograms of the {\it TESS} data (Fig.\,\ref{spTESS}) when aliasing is accounted for. The Fourier analysis suggest that any rotational modulation would instead occur with a period not too far away from the orbital period. Hence, it seems rather difficult to explain the peculiarities of BD+60$^{\circ}$\,497 by pseudo-synchronised rotation of a spotted secondary star.

\subsection{The possibility of a triple system}
We consider the possibility that BD+60$^{\circ}$\,497 might be a hierarchical triple system formed by the inner close binary and a third star on a wider orbit. Such a scenario could explain some of the astrometric properties of the system. Indeed, \citet{Bai21} derived a somewhat larger distance from the {\it Gaia} EDR3 parallaxes than for other O-type stars of IC\,1805, and \citet{Hun24} did not consider BD+60$^{\circ}$\,497 to be a member of IC\,1805. If this conclusion is confirmed by future {\it Gaia} data releases, BD+60$^{\circ}$\,497 could then be part of the Cas OB6 association but not of the IC\,1805 cluster. Taken at face value, the distance inferred by \citet{Bai21} would leave some room in the derived absolute magnitude for a third light contribution by a tertiary star (see Sect.\,\ref{recspec}). Moreover, the proper motion in right ascension of BD+60$^{\circ}$\,497 also deviates from that of most other O-type stars in IC\,1805, though it remains in the range of OB stars of the cluster (see Appendix\,\ref{AppB}). Such deviations could reflect the motion in a triple system.

We investigated whether such a scenario could explain the spectroscopic and photometric oddities found in our analysis. Concerning the epoch dependence of the reconstructed individual spectra, the spectral signatures of a third star would definitely bias the results of the disentangling procedure, which assumes the contribution of only two stars. However, there is no obvious reason why this should preferentially affect the reconstructed secondary spectrum. Given the mass ratio we derived, the contamination should impact the reconstructed spectra of both stars nearly equally.

As to the light curves, the presence of a tertiary star would not inhibit the proximity effects expected for the close binary, but the third light would dilute their signatures in the global photometry of the triple system. If the inclination of the outer orbit is close to $90^{\circ}$, one could also expect eclipses of the third star by the components of the inner binary. However, given the low level of the photometric variations, such events would have to be grazing eclipses. The changing configuration of the binary, due to apsidal motion, could then explain the sector-to-sector changes, but the duration of such grazing eclipses should be rather short, unlike the minima seen in the light curves which actually last at least 0.25 in phase.  

In conclusion, whilst we cannot rule out the possibility of a triple system, such a scenario does not seem to easily explain the oddities of the BD+60$^{\circ}$\,497 system. The forthcoming {\it Gaia} data releases should shed new light on the relationship between this system and the IC\,1805 cluster, and hopefully give further constraints on the possible presence of a tertiary component.

\subsection{Comparison with other massive binaries}
           Finally, we briefly compare the light curve of BD+60$^{\circ}$\,497 with those of other eccentric, short-period, non-eclipsing, massive binaries. \citet{She22} present light curves for O-type binaries in the Tarantula Nebula. Among the non-eclipsing binaries in their sample, VFTS\,475 (O9.7\,V + B0\,V, $P_{\rm orb} = 4.05$\,d, $e = 0.573$) and VFTS\,887 (O9.7\,V + O9.5\,V, $P_{\rm orb} = 2.67$\,d, $e = 0.056$) are those that most closely ressemble BD+60$^{\circ}$\,497. Their light curves display peak-to-peak amplitudes of $\leq 10$\,mmag (VFTS\,887) and $\sim 15$\,mmag (VFTS\,475), which are close to the peak-to-peak amplitude of the variations seen in BD+60$^{\circ}$\,497. Unfortunately, \citet{She22} did not present a modelling of these light curves, preventing a more detailed comparison.
             
             Among the well-studied O-type binaries of our Galaxy, HD\,165052 is probably the best case to compare with BD+60$^{\circ}$\,497. HD\,165052 is an O6.5\,V((f)) + O7\,V((f)) system with an eccentric, short-period, orbit \citep[$e = 0.098$, $P_{\rm orb} = 2.95585$\,d;][and references therein]{Ros23}. We analysed the {\it TESS} photometry of this non-eclipsing system (see Appendix\,\ref{AppG}). As for BD+60$^{\circ}$\,497, we found variability at the 10\,mmag level, part of which could not be reproduced with a binary model assuming the system parameters found in the spectroscopic study.
             
             It thus seems that the difficulties to reproduce the light curves of short-period, non-eclipsing, eccentric, massive binaries with binary models are not restricted to the sole case of BD+60$^{\circ}$\,497. This might indicate that current models lack some ingredients in the description of such binaries. Whilst an in-depth investigation of the possible causes is beyond the scope of the present work, a promising candidate could be the effect of external radiation pressure on the shape of the stars \citep{Pal13}. External radiation pressure, that is due to illumination by a luminous companion, acts against the gravitational pull of the companion over the illuminated part. Compared to models without radiation pressure, the drop-shaped deformation of the stellar surface facing the $L_1$ point is attenuated when radiation pressure is accounted for \citep[see Fig.\,3 of][]{Pal13}. 

\section{Summary and conclusions}
We have presented the first analysis of the RVs and {\it TESS} photometry of the close eccentric O-star binary system BD+60$^{\circ}$\,497 accounting for the presence of apsidal motion. This study enabled us to determine a total rate of apsidal motion of $\dot{\omega} = (6.15^{+1.05}_{-1.65})^{\circ}$\,yr$^{-1}$. When comparing with the theoretical expectations from the stellar structure models of \citet{Cla19}, this led to an age estimate of the system of $4.13^{+0.42}_{-1.37}$\,Myr. This estimated value is in very reasonable agreement with the age of the IC\,1805 cluster (3.5\,Myr) found by \citet{Sun17}.

However, our analysis also revealed a number of oddities that challenge our understanding of this system. First, we found a strange change in the reconstructed secondary spectrum of the epochs 2002-2003 and 2018-2022. When comparing with standard spectral classification criteria, the secondary had its spectral type change from O8.5 in 2002-2003 to O7 in 2018-2022. Also, the relative optical brightness of the secondary would have increased from 35\% to 50\%. Such a radical change is difficult to explain. It is most probably related to the apparent changes in the visibility of the secondary lines with orbital phase, which were previously observed by \citet{Rau04}. This could hint at a non-uniform surface temperature distribution. We investigated several possible origins of such a feature (mutual heating, back-warming by radiation from a wind interaction zone), but it is currently unclear whether these effects can explain the observations.

The second challenging result of our study concerns the light curves, which display a single broad minimum and changes in the overall shape. Proximity effects, especially the tidal deformation of the stars, are expected to produce a light curve with two minima around conjunction phases. Not only do the observed light curves lack the second minimum, but the single broad minimum appears offset in orbital phase compared to the expectations. We investigated a number of possible explanations (action of TEOs, wind effects, magnetic spots), but none of them provides an obvious solution to the problem, though some (especially TEOs and magnetic spots) produce effects that go in the right direction.

To date, BD+60$^{\circ}$\,497 remains a challenging object for our understanding of proximity effects in massive binaries. Future spectroscopic observations will be extremely important to monitor the behaviour of the secondary spectrum and will hopefully provide a better understanding of the causes of its unusual behaviour. Such observations will further allow additional RV data to be accumulated, thereby extending the time frame over which apsidal motion can be studied. Another important aspect are the oddities of the light curve, especially since similar effects seem to be present in other non-eclipsing short-period eccentric massive binaries, such as HD\,165052. This could imply that current models for proximity effects in such systems are too simplistic.   

\begin{acknowledgements}
YN acknowledges support from the Fonds National de la Recherche Scientiﬁque (Belgium). This research was supported by the University of Li\`ege under the Special Funds for Research, IPD-STEMA Programme. This work uses data collected by the {\it TESS} mission, which are publicly available from the Mikulski Archive for Space Telescopes (MAST). Funding for the TESS mission is provided by NASA’s Science Mission directorate. ADS and CDS were used during this research.
\end{acknowledgements}

\bibliographystyle{aa}
\bibliography{mybiblio}

\appendix
\section{Table of new RV data}
Table\,\ref{Journal} presents the heliocentric RVs determined via spectral disentangling applied to the Aur\'elie and HEROS spectra of BD+60$^{\circ}$\,497. These RVs were used to determine the rate of apsidal motion. Typical errors, estimated from the dispersion of the RVs obtained via disentangling over three different wavelength domains, are 10\,km\,s$^{-1}$ for the primary and 15\,km\,s$^{-1}$ for the secondary star.
\begin{table}[h!]
  \caption{Heliocentric RVs of BD+60$^{\circ}$\,497 determined via spectral disentangling.\label{Journal}}
  \begin{center}
  \begin{tabular}{c c r r}
    \hline
    Date & Inst. & RV$_p$ & RV$_s$ \\
    HJD-2450000 & & (km\,s$^{-1}$) & (km\,s$^{-1}$) \\
    \hline
2520.644 & T152  & $-115.3$  & $ 116.8$ \\
2523.600 & T152  & $  92.0$  & $-219.7$ \\
2524.552 & T152  & $-111.4$  & $ 108.2$ \\
2527.555 & T152  & $  91.9$  & $-216.7$ \\
2528.533 & T152  & $-122.8$  & $ 103.6$ \\
2529.562 & T152  & $-152.5$  & $  64.6$ \\
2531.543 & T152  & $  87.0$  & $-219.2$ \\
2532.534 & T152  & $-135.9$  & $ 105.9$ \\
2916.583 & T152  & $-156.2$  & $ 119.8$ \\
2918.667 & T152  & $  61.3$  & $-194.5$ \\
2919.632 & T152  & $  55.5$  & $-211.8$ \\
2922.677 & T152  & $  73.1$  & $-184.9$ \\
2928.641 & T152  & $-171.0$  & $ 148.3$ \\
2934.545 & T152  & $  69.4$  & $-186.3$ \\
6682.573 & TIGRE  & $-202.8$  & $  62.3$ \\
6684.571 & TIGRE  & $  86.0$  & $-203.4$ \\
6686.579 & TIGRE  & $-202.4$  & $  53.2$ \\
6688.573 & TIGRE  & $  82.3$  & $-208.8$ \\
8153.613 & TIGRE$^*$ & $  44.6$  & $-198.3$ \\
8156.587 & TIGRE$^*$ & $  62.9$  & $-134.6$ \\
8165.578 & TIGRE$^*$ & $  45.5$  & $-195.7$ \\
8170.590 & TIGRE$^*$ & $-178.5$  & $  61.0$ \\
8172.591 & TIGRE$^*$ & $  54.0$  & $-172.3$ \\
8174.588 & TIGRE$^*$ & $-173.3$  & $  69.8$ \\
8353.542 & T152  & $-161.2$  & $  93.9$ \\
8353.563 & T152  & $-158.2$  & $  94.1$ \\
8354.547 & T152  & $  75.1$  & $ -88.7$ \\
8354.569 & T152  & $  75.9$  & $ -86.8$ \\
8356.544 & T152  & $-154.6$  & $   6.2$ \\
8356.572 & T152  & $-170.6$  & $  17.4$ \\
8357.520 & T152  & $-161.2$  & $  91.6$ \\
8357.541 & T152  & $-155.0$  & $  89.5$ \\
8719.586 & T152  & $  68.1$  & $-180.6$ \\
8719.614 & T152  & $  68.3$  & $-179.8$ \\
8721.550 & T152  & $-211.4$  & $ 103.7$ \\
8721.578 & T152  & $-196.2$  & $ 111.0$ \\
9851.603 & T152  & $  87.1$  & $-192.1$ \\
9851.630 & T152  & $  88.5$  & $-197.4$ \\
9851.660 & T152  & $  89.3$  & $-198.6$ \\
9853.551 & T152  & $-173.2$  & $ 139.9$ \\
9853.581 & T152  & $-167.0$  & $ 146.6$ \\
9853.637 & T152  & $-181.1$  & $ 139.8$ \\
\hline
  \end{tabular}
  \end{center}
  \tablefoot{The second column specifies the instrument used: T152 refers to the 1.52m telescope at OHP equipped with the Aur\'elie spectrograph, whilst TIGRE was equipped with the HEROS spectrograph. For the TIGRE spectra marked with an asterisk ($^*$) only red channel data were available.}
\end{table}

\section{Uncertainties on the orbital solution\label{Appnew}}
\begin{figure}[h]
  \includegraphics[angle=0, width=9cm]{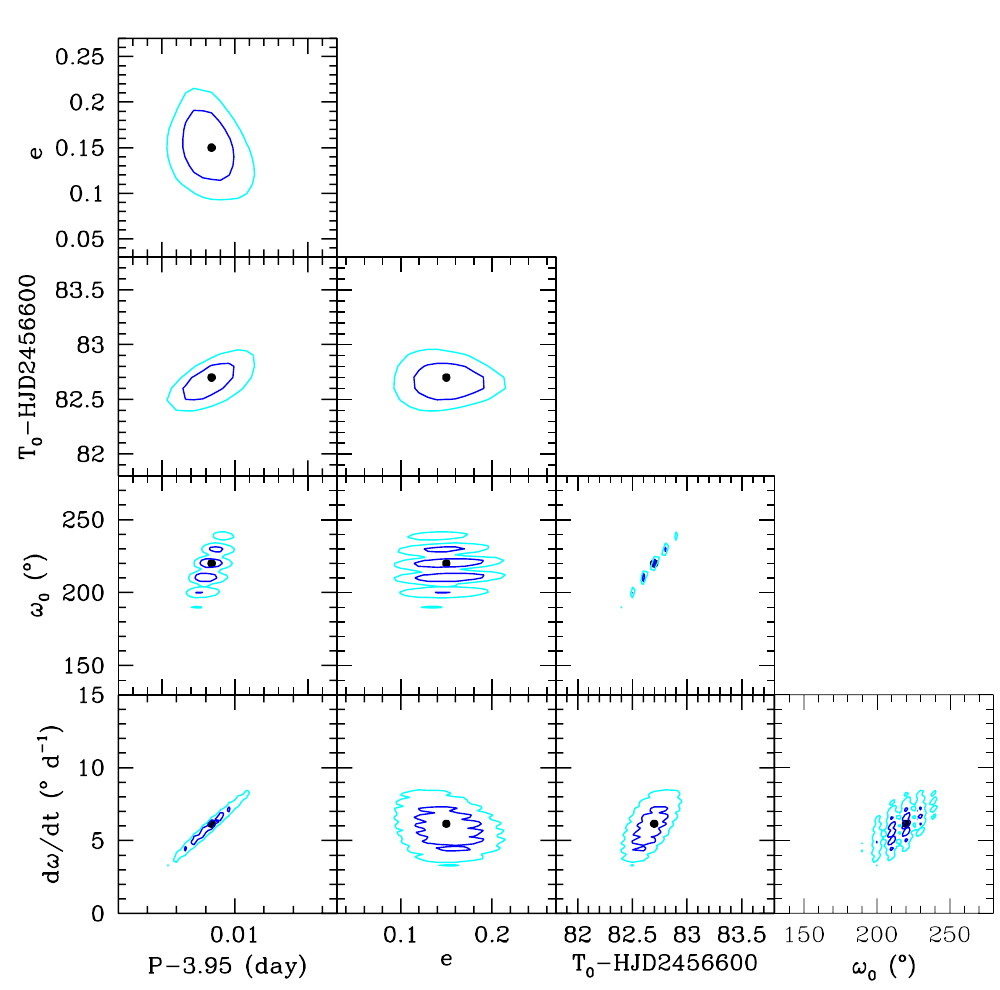}
  \caption{Corner plot for the five parameters ($P_{\rm anom}$, $e$, $t_0$, $\omega_0$ and $\dot{\omega}$) for $K_p = 153.5$\,km\,s$^{-1}$ and dataset 2. The black dots indicate our best solution (see Table\,\ref{omegadotfit}), whilst the blue and cyan contours correspond respectively to $1\,\sigma$ and 99\% confidence levels.\label{cornerplot}}
\end{figure}

        For a given value of $K_p$, the best-fit orbital solution accounting for apsidal motion was searched for in a five-dimensional parameter space by minimising the $\chi^2$ between the observed and computed primary RVs. To assess the uncertainties on the various parameters, we computed the $\chi^2$ values corresponding to a $1\,\sigma$ and 99\% confidence level. The associated iso-$\chi^2$ surfaces in the five-dimensional space were then projected onto all two-dimensional planes, that is all combinations of two parameters that can be defined. Figure\,\ref{cornerplot} illustrates the corresponding contours for the $K_p = 153.5$\,km\,s$^{-1}$ case and for the RV dataset 2. These contours allow the $1\,\sigma$ uncertainties quoted in Table\,\ref{omegadotfit} to be established. They also highlight correlations between $\dot{\omega}$ and $P_{\rm anom}$ on the one hand and between $\omega_0$ and $t_0$ on the other.

\section{Astrometric parameters \label{AppB}}
\begin{figure}[t!]
    \includegraphics[angle=0, width=7.8cm]{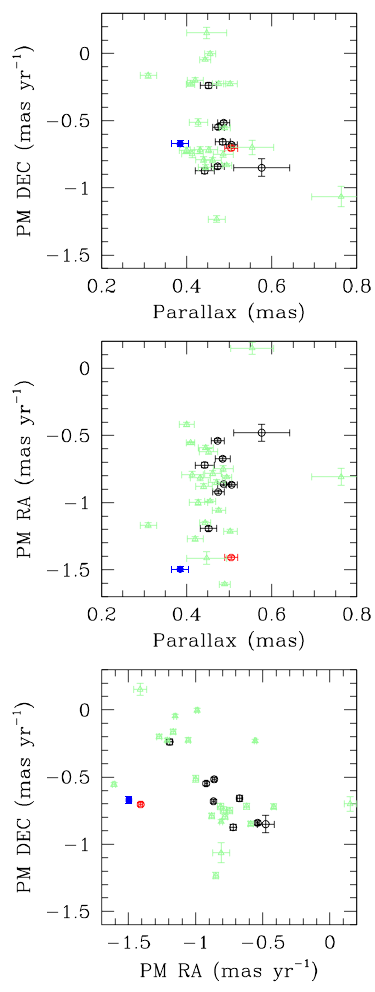}
  \caption{Gaia astrometric parameters of OB stars within a radius of 30\arcmin\ around the centre of the IC\,1805 open cluster. The blue symbols stand for BD+60$^{\circ}$\,497. Black open circles indicate O-type star, whilst the green triangles stand for B0-B3 stars. The red circle stands for the O9.5 star KM\,Cas whose membership in the cluster was questioned.\label{Gaiaplot}}
\end{figure}

Figure\,\ref{Gaiaplot} displays the {\it Gaia} DR3 astrometric parameters of BD+60$^{\circ}$\,497 (blue symbols) compared to those of OB stars within a radius of 30\arcmin\ of the centre of IC\,1805. The OB stars were selected using the TGAS catalogue\footnote{https://vizier.cds.unistra.fr/viz-bin/VizieR-3?-source=I/337/tgas} \citep{Gai16} restricted to those objects with spectral types up to B3 in the SIMBAD\footnote{https://simbad.u-strasbg.fr/simbad/} database. This sample includes eight O-type stars thought to belong to IC\,1805 \citep[][HD\,15558, HD\,15570, HD\,15629, BD+60$^{\circ}$\,498, BD+60$^{\circ}$\,499, BD+60$^{\circ}$\,501, BD+60$^{\circ}$\,513, and IC\,1805 188, shown by black open circles]{Jos83}, one O9.5 star (KM\,Cas, open red circle) whose membership in IC\,1805 was questioned by \citet{Sun17}, and 21 B-type stars with spectral types earlier than B3 (shown by green triangles). All objects with large error bars on their astrometric parameters do have a RUWE that strongly exceeds the usual limit of 1.4. For instance, the only O-star system with very big error bars is the long-period binary HD\,15558 \citep{Rau16} for which the astrometric solution has a RUWE of 3.06.

Based on proper motions from the {\it Gaia} DR1 \citep{Lin16}, \citet{Sun17} concluded that the proper motions of B-type stars in IC\,1805 displays a large scatter and follows an elongated distribution in the proper motion plane. With the {\it Gaia} DR3 data, the scatter is considerably reduced compared to Fig.\,32 of \citet{Sun17}, though we still find a somewhat elongated distribution. BD+60$^{\circ}$\,497 lies a bit off the main locus of the O-type stars both in proper motion and parallax, but remains within the region occupied by the early B-type stars. A definite answer regarding the membership of BD+60$^{\circ}$\,497 in the IC\,1805 cluster will have to await a full re-analysis of the astrometry of IC\,1805 based on the forthcoming {\it Gaia} Data Release 4 (due in December 2026).
\section{{\it Nightfall} fits of Sectors 85 and 86 \label{AppC}}
\begin{figure}[h]
    \includegraphics[angle=0, width=7.8cm]{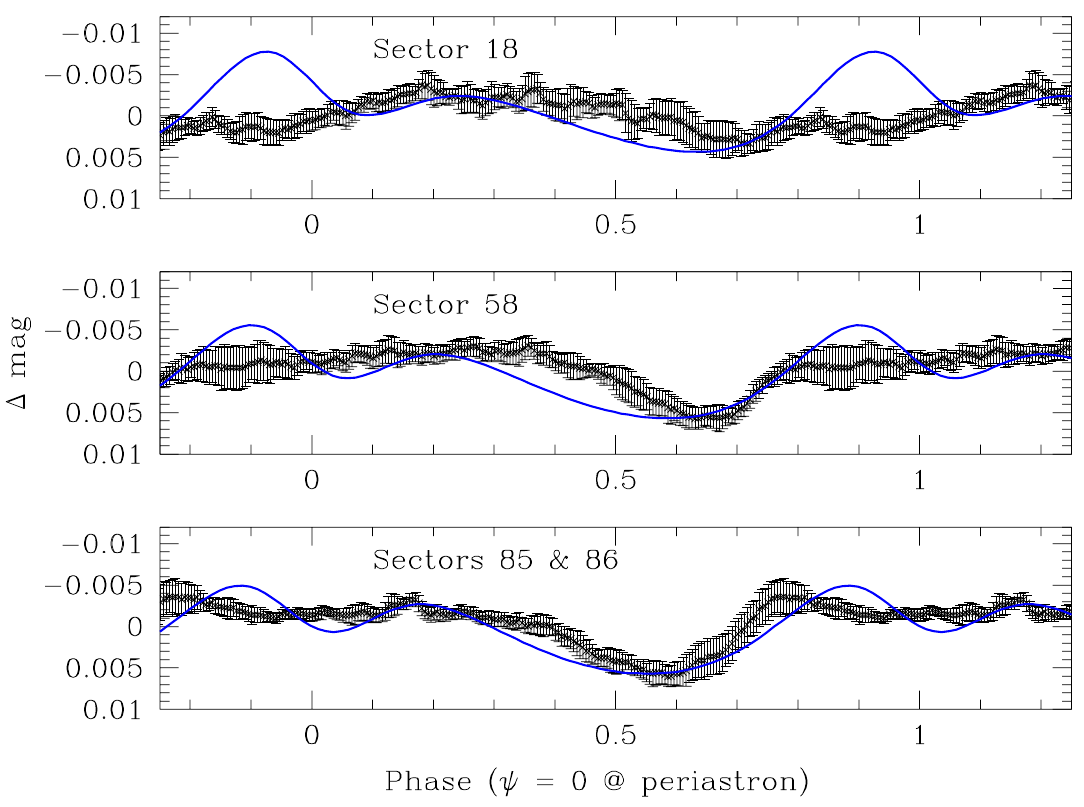}
  \caption{Phase-folded {\it TESS} light curves of BD+60$^{\circ}$\,497. The blue curve shows the {\it Nightfall} model that offers a reasonable agreement with the data from Sectors 85 and 86 (bottom panel). Phase 0.0 corresponds to periastron passage. The top and middle panel illustrate the same model with the $\omega$ values propagated backwards to Sectors 18 and 58 (see text).\label{appendLC}}
\end{figure}

Figure\,\ref{appendLC} displays the {\it TESS} light curves compared to {\it Nightfall} models accounting for proximity effects and designed to adjust as well as possible the data from Sectors 85 and 86. The model parameters are $q = 1.28$, $i = 45^{\circ}$, $fill_p = fill_s = 0.6$, $T_{{\rm eff}, p} = 37.5$\,kK, $T_{{\rm eff}, s} = 32.5$\,kK, $e = 0.15$, and $\omega = 255^{\circ}$. We note here that the $\omega$ value was left as a free parameter, and that its value does not agree with our apsidal motion solution. The Roche lobe filling factors correspond to mean stellar radii of 7.3\,R$_{\odot}$ for the primary and 6.5\,R$_{\odot}$ for the secondary, which are significantly lower than the radii from the evolutionary models of \citet{Cla19} - 9.8\,R$_{\odot}$ for the primary and 7.3\,R$_{\odot}$ for the secondary. The secondary temperature corresponds to the lower value found by comparing our disentangled spectra with synthetic spectra. This relatively low value is required to keep the amplitude of the predicted photometric minimum near periastron at a low level. In this figure, we also show the light curves for Sectors 18 and 58 with the same parameters, except for $\omega$, which we took to be respectively $232^{\circ}$ and $246^{\circ}$. The data from these sectors are clearly not well fitted. 

\section{Additional CoMBiSpeC simulations}
\begin{figure*}[t!]
  \begin{minipage}{8.5cm}
    \includegraphics[angle=0, width=8.5cm]{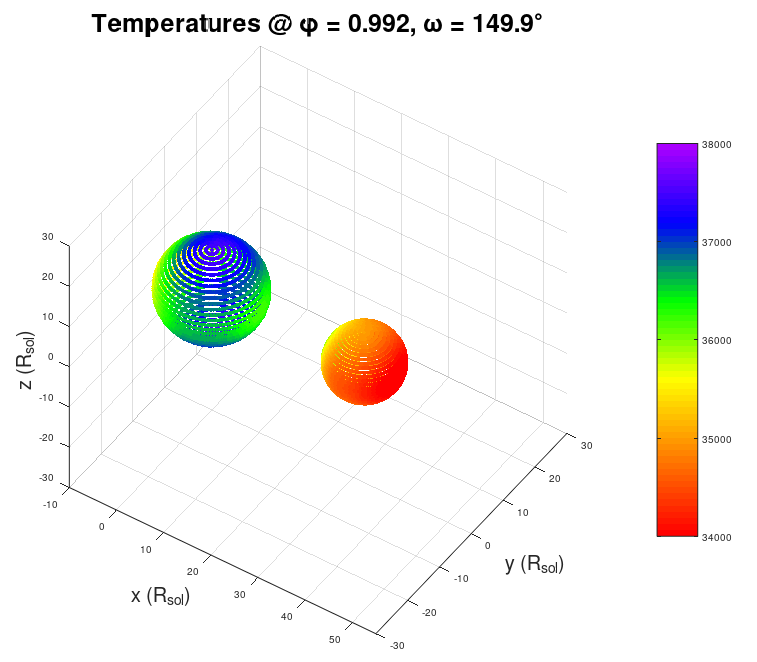}
  \end{minipage}
  \hfill
  \begin{minipage}{8.5cm}
    \includegraphics[angle=0, width=8.5cm]{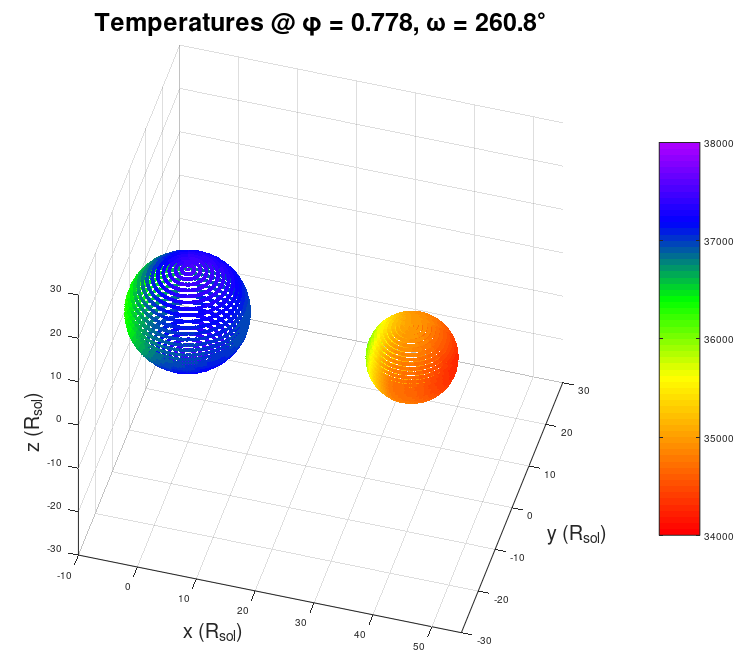}
  \end{minipage}
  \begin{minipage}{8.5cm}
    \includegraphics[angle=0, width=8.5cm]{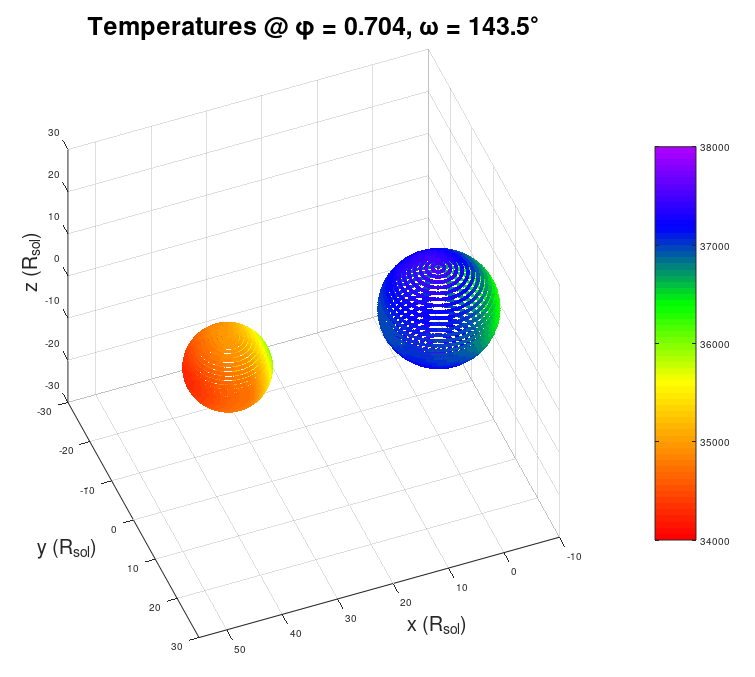}
  \end{minipage}
  \hfill
  \begin{minipage}{8.5cm}
    \includegraphics[angle=0, width=8.5cm]{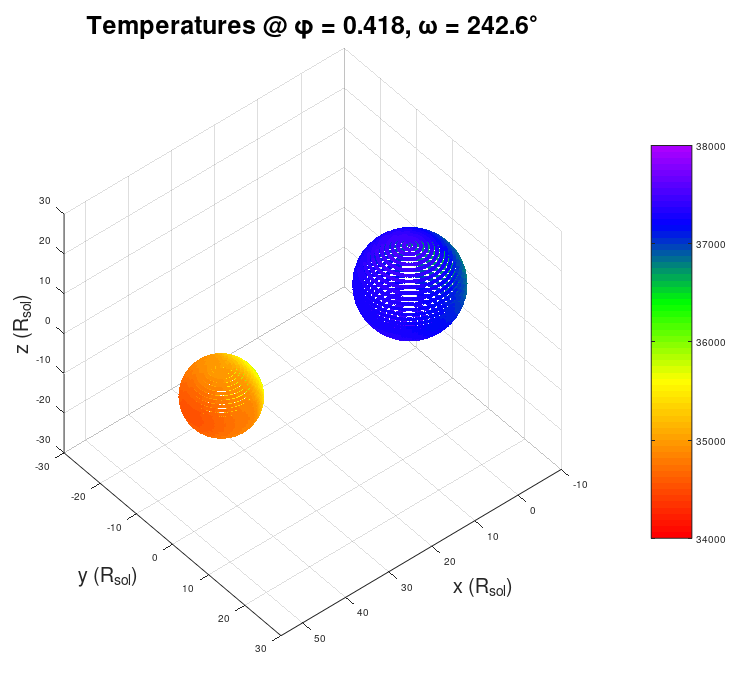}
  \end{minipage}  
  \caption{Surface temperature distribution computed with CoMBiSpeC for the stars of the BD+60$^{\circ}$\,497 binary system at the phases corresponding to the spectra in Fig.\,\ref{montage}. The stars are displayed as they would be seen from Earth, assuming an orbital inclination of $45^{\circ}$. The top left (respectively top right) panel stands for the configuration corresponding to the 2002-2003 (respectively 2018-2022) epoch spectrum in the top row of Fig.\,\ref{montage}. The bottom left and bottom right panels likewise correspond to the two epochs in the bottom row of Fig.\,\ref{montage}.\label{mutual}}
\end{figure*}

Figure\,\ref{mutual} illustrates the surface temperature distributions computed with CoMBiSpeC for the orbital configurations corresponding to the spectra displayed in Fig.\,\ref{montage}.

\section{Simulations in MESA}\label{app:mesa-input}
In our simulations, we used MESA version 24.08.1, compiled with the MESA Software Development Kit \citep[SDK version 24.7.1;][]{Tow24}. The orbital parameters of BD+60$^\circ$\,497 and physical parameters of its components were the same as those used in our CoMBiSpeC simulations (Sect.\,\ref{sect:combispec}).

We assumed that both components had identical initial chemical composition with a solar-scaled mixture of elements according to \citet{Asp09}, and the corresponding opacity tables for this mixture. Our models are characterised by the initial mass fraction of helium, $Y=0.28$, and metallicity, $Z=0.02$. We assumed both stars rotate rigidly. We tested many combinations of the rotation rates of the components, experimenting with values that were in the range of 50\% to 200\% of the rotation rate estimated in our paper. MESA uses the mathematical formalism of \citet{Heg00} and \citet{Heg05} to apply structural corrections, perform different types of rotationally induced mixing and `diffusion' of angular momentum between adjacent shells. We used the convective pre-mixing scheme \citep[][their Sect.~5]{Pax19} in combination with the Ledoux criterion to define the boundaries of convective instability. Convective mixing was incorporated into the models via mixing length theory in the formalism provided by \citet{Hen65}, with the value of the solar-calibrated mixing length parameter $\alpha_{\rm MLT}=1.82$ \citep{Cho16}. We ignored the effects of semiconvection and thermohaline mixing. The overshooting of the material above the convective core was treated in the exponential diffusion approximation developed by \citet{Her00} with an adjustable parameter, $f_{\rm ov}=0.015$. Mass loss rate due to the radiation-driven stellar wind has been calculated according to the prescription of \citet{Bjo23}. During the integration, we allowed for the circularisation of the orbit and synchronisation of the rotation rate of the components, following the formulae incorporated in the MESA {\tt binary} package for stars with radiative envelopes \citep{Pax15}. For detailed information on all the above processes and their implementation in the MESA code, we refer the reader to the series of MESA instrumental papers.

\section{{\it TESS} light curve of HD\,165052 \label{AppG}}
HD\,165052 is an eccentric ($e = 0.098$), non-eclipsing, short-period ($P_{\rm orb} = 2.95585$\,d), O6.5\,V((f)) + O7\,V((f)) binary system displaying a relatively fast apsidal motion \citep[$\dot{\omega} = 11.3^{\circ}$\,yr$^{-1}$,][]{Ros23}. The {\it TESS} satellite observed this system in full frame mode, with a 200\,s cadence, during two consecutive sectors (91 and 92, April - May 2025). We retrieved the full-frame images (FFI) data from the MAST portal and extracted the light curves with the Python software package Lightkurve. Aperture photometry was extracted on $51 \times 51$ pixels image cutouts. For the source mask, we adopted a flux threshold of 15 times the median absolute deviation over the median flux. The background was evaluated from those pixels of the image cutouts that were below the median flux. Background subtraction was performed by means of a principal component analysis including five components. Whilst the {\it Gaia} DR3 catalogue lists over 300 sources within a 1\,arcmin radius, only one of these sources has a G-magnitude brighter than 13. Even this source is 6 magnitudes fainter than HD\,165052. We thus conclude that there is no significant contamination of the {\it TESS} photometry by neighbouring sources.

\begin{figure}
    \includegraphics[angle=0, width=8.5cm]{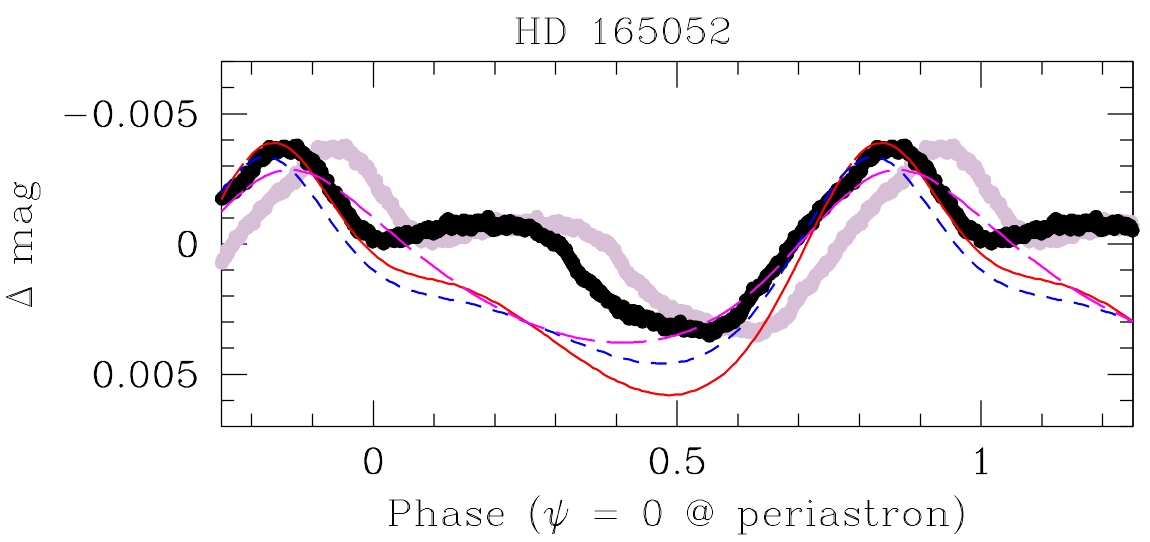}
  \caption{Phase-folded {\it TESS} light curve of the HD\,165052 binary system as observed during sectors 91 and 92 (grey symbols), and shifted in phase by $-0.08$ (black symbols). The mean error per phase-binned data point is $1.6\,10^{-4}$\,mag. The red continuous line illustrates the synthetic curve computed with the set of parameters taken from \citet{Ros23}. The magenta long-dashed curve shows the best-fit synthetic curve with the inclination left as a free parameter, whilst the blue short-dashed curve corresponds to the best-fit for an inclination of $i = 22.7^{\circ}$. The model parameters are given in Table\,\ref{Tab165052}.\label{HD165052}}
\end{figure}

The 17169 photometric data points from both sectors were combined and phase-folded using the ephemerides of \citet{Ros23}. The resulting light curve, binned into 200 phase bins (see the grey points in Fig.\,\ref{HD165052}), displays a well-defined peak shortly before periastron. The brightness than decreases and reaches a plateau, until it drops towards a broad minimum centred on apastron. If we compute a synthetic light curve with the {\it Nightfall} code, adopting the system and stellar parameters from \citet{Ros23}, we find some important discrepancies. First of all, we find a shift by $-0.080$ in phase between the synthetic curve and the observed one. Such a shift could be explained by the $5 \times 10^{-5}$\,d uncertainty on $P_{\rm orb} = 2.95585$\,d propagated over the 15 years between the time of periastron given by \citet{Ros23} and the epoch of the {\it TESS} data. We thus applied this shift to the data (Fig.\,\ref{HD165052}) and started a fitting procedure with {\it Nightfall}. Since the predicted amplitude of the variations exceeded the observed value, we let the Roche lobe filling factors, the orbital inclination and $\omega$ vary in the fitting procedure. This resulted in a very low orbital inclination ($i = 11.4^{\circ}$), which would imply unrealistically large masses of 180 and 162\,M$_{\odot}$ for the primary and secondary, respectively. The best-fit model still failed to reproduce the plateau-like part of the observed light curve between phases 0.0 and 0.35.  

In a second attempt, we froze the inclination to $i = 22.7^{\circ}$ \citep{Ros23}, leaving only the filling factors and $\omega$ as free parameters. The resulting best-fit adjustment was of lower quality than in the previous case (see Table\,\ref{Tab165052}). It failed again to reproduce the plateau in the light curve. Moreover, the best-fit Roche-lobe filling factors were smaller than those assumed by \citet{Ros23}, which, given the $(R/a)^5$ dependence of $\dot{\omega}$, would imply a lower than observed rate of apsidal motion.  

In summary, we find that the observed photometric variations of HD\,165052 cannot easily be explained with the binary model assuming the parameters inferred from the spectroscopic study of \citet{Ros23}, or leaving some of them free. This situation is very much reminiscent of the problems we encountered for BD+60$^{\circ}$\,497. Unfortunately, there is no multi-epoch high-precision photometry available for HD\,165052. Therefore, we cannot check whether or not the observed light curve displays variations on timescales of years as we found for BD+60$^{\circ}$\,497.

\begin{table}
  \caption{Parameters used in the modelling of the {\it TESS} light curve of HD\,165052.\label{Tab165052}}
  \begin{center}
  \begin{tabular}{c c c c}
    \hline
    \multicolumn{4}{c}{Fixed parameter values} \\
    \hline
    $t_0$ (HJD) & \multicolumn{3}{c}{2455050.00} \\
    $P_{\rm anom}$ (d) & \multicolumn{3}{c}{2.95585$^{\dagger}$} \\
    $e$ & \multicolumn{3}{c}{0.098} \\
    $m_p/m_s$ & \multicolumn{3}{c}{1.109} \\
    $T_p$ (kK) & \multicolumn{3}{c}{37.5} \\
    $T_s$ (kK) & \multicolumn{3}{c}{36.0} \\
    \hline
    Parameter & Default & $i$ free & $i$ frozen \\
    \hline
    $\omega$ ($^{\circ}$) & 227 & $219.0 \pm 1.3$ & $215.7 \pm 2.1$\\
    $i$ ($^{\circ}$) & 22.7 & $11.5 \pm 0.8$ & 22.7 \\
    $fill_p$ & 0.646 & $0.546 \pm 0.006$ & $0.622 \pm 0.011$ \\
    $fill_s$ & 0.600 & $0.567 \pm 0.010$ & $0.447 \pm 0.046$ \\
    $\chi^2$ & 146.9 & 86.9 & 120.1 \\    
\hline
  \end{tabular}
  \end{center}
  \tablefoot{$^{\dagger}$ The phase of the observed light curve, computed with these ephemerides, was shifted by $-0.080$ to bring it into agreement with the model. This corresponds to the propagation of the error on $P_{\rm anom}$ from $t_0$ to the epoch of the {\it TESS} observations.\newline The bottom part of the table shows the parameters of the curves in Fig.\,\ref{HD165052}. The `Default' column refers to the red curve computed with the parameters taken from \citet{Ros23}. The `$i$ free' and `$i$ fixed' columns correspond respectively to the magenta long-dashed and blue short-dashed curves.}
\end{table}

\end{document}